\documentclass[reprint, aps,prfluids, amsmath,amssymb]{revtex4-2}
\usepackage{graphicx}
\usepackage{dcolumn}
\usepackage{bm}

\usepackage[utf8]{inputenc}
\usepackage[T1]{fontenc}
\usepackage{mathptmx}
\usepackage{etoolbox}
\usepackage{xcolor}
\usepackage{hyperref}
\usepackage[export]{adjustbox}

\begin{document}

\preprint{APS/123-QED}

\title{Multi-fidelity validation of variable-density turbulent mixing models}


\author{Benjamin Musci}%
 \email{bmusci3@gatech.edu}
\altaffiliation[Also at ]{Lawrence Livermore National Laboratory} 
\affiliation{Georgia Institute of Technology
}

\author{Britton Olson}%
 \email{olson45@llnl.gov}
\affiliation{ 
Lawrence Livermore National Laboratory 
}%

\author{Samuel Petter}
\affiliation{Georgia Institute of Technology \\
}

\author{Gokul Pathikonda}
\altaffiliation[Also at ]{Arizona State University}
\affiliation{Georgia Institute of Technology \\
}

\author{Devesh Ranjan}
 \email{devesh.ranjan@me.gatech.edu}
\affiliation{Georgia Institute of Technology \\
}

\date{\today}

\begin{abstract}

In this study, ensembles of experimental data are presented and utilized to compare and validate two models used in the simulation of variable density, compressible turbulent mixing.  Though models of this kind (Reynolds Averaged Navier-Stokes and Large-Eddy Simulations) have been validated extensively with more canonical flows in previous studies, the present approach offers novelty in the complexity of the geometry, the ensemble based validation, and the uniformity of the computational framework on which the models are tested.  Moreover, all experimental and computational tasks were completed by the authors which has led to a tightly coupled experimental configuration with its "digital twin.” The experimental divergent-shock-tube facility and its data acquisition methods are described and replicated in simulation space.  A 2D Euler model which neglects the turbulent mixing at the interface is optimized to experimental data using a Gaussian process.  This model then serves as the basis for both the 2D RANS and 3D LES studies that make comparisons to the mixing layer data from the experiment.  RANS is shown to produce good agreement with experimental data only at late flow development times. The LES ensembles generally show good agreement with experimental data, but display sensitivity to the characterization of initial conditions. Resolution dependent behavior is also observed for certain higher-order statistics of interest. Overall, the LES model successfully captures the effects of divergent geometry, compressibility, and combined non-linear instabilities inherent to the problem. The successful prediction of mixing width and its growth rate highlight the existence of three distinct regimes in the development of the instability, each with similarities to previously studied instabilities. 

\end{abstract}

\maketitle


\newcommand{\rev}[1]{\textcolor{black}{#1}}
\newcommand{\revtwo}[1]{\textcolor{black}{#1}}
\newcommand{\brit}[1]{\textcolor{red}{#1}}
\newcommand{\orderof}{\ensuremath{\mathcal{0}}}
\section{Introduction} \label{sec:intro}

The mixing of variable-density flows occurs across a vast range of space and time scales. It occurs on Astrophysical scales in events like Supernovae (SN), on kilometer scales in oceanic and atmospheric flows, all the way down to the millimeter scales of Inertial Confinement Fusion (ICF). In ICF and SN in particular, it is particularity challenging to recreate the exact mixing for scientific observation and study. This difficulty in physical observation has placed a great need for the use of computer simulations as tools for increasing understanding. However, these powerful models can in turn create more questions, and as such their efficacy requires experimental validation. For instance, a simulation used to model the multiple combined instabilities occurring in an ICF capsule should first be shown to successfully model the mixing driven by a single combined instability in a meter-scale experiment. The current work aims to achieve just that: use high-fidelity, combined-instability experimental data to create and validate a "digital-twin" simulation using two common turbulent mixing-models.

\subsection{\label{subsec:problem}Problem Outline and Motivation}

\begin{figure*}[ht!]
\includegraphics[scale=0.84]{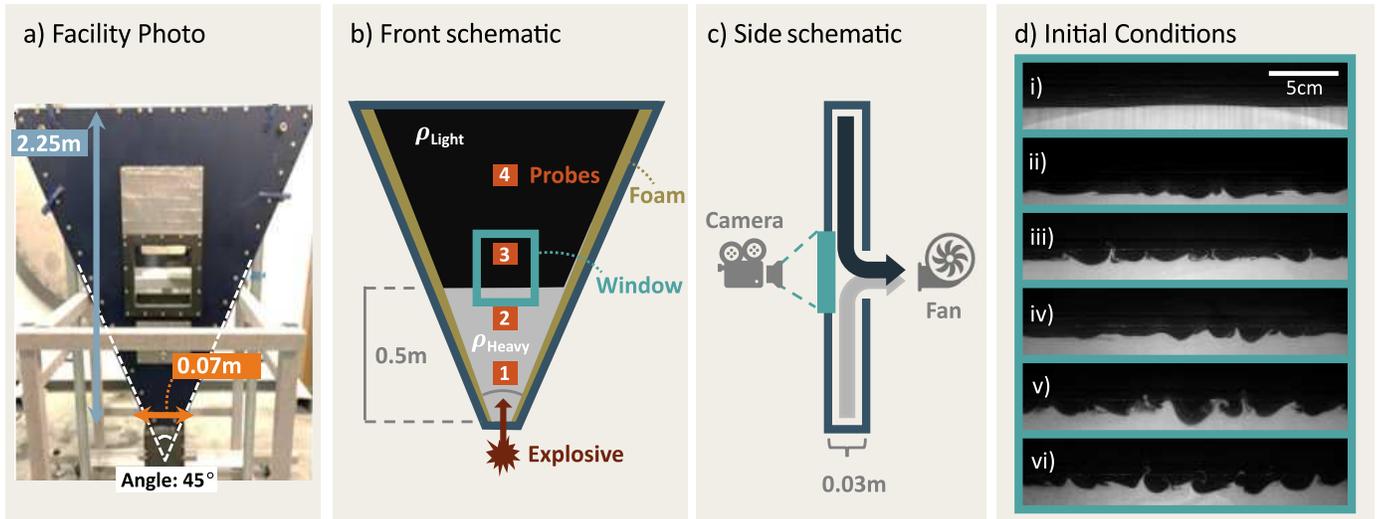}
\caption{\label{fig:facility} Pictures and diagrams of the experimental facility. a) Front facing picture of the diverging facility with key dimensions annotated. The viewing window can be seen in the center of the image. b) Front facing schematic of the facility. The location of the foam layer, viewing window, and commercial detonator are all indicated. The heavy and light gas are also indicated and colored according to their appearance in Mie Scattering images. Note that for PIV, the data come from a smaller field of view located within the window indicated here.  c) Side facing schematic of the facility showing facility width, gas entry and exit points, as well as fan and camera locations. d) Individual initial condition realizations visualized using Mie Scattering. i: Flat interface condition. The incident blast wave can also be seen (brightly shaded region) impacting the interface. ii-vi: An array of perturbed initial conditions demonstrating the range of different perturbation amplitudes and wavelengths generated by the facility. }
\end{figure*}

The interaction of a blast wave (BW) and a multi-component gaseous interface results in a combined hydrodynamic instability when the BW travels from a high to low density gas \cite{Arnett1989, Muller1991}. In this case, the blast-interface interaction incites the Richtmyer-Meshkov Instability (RMI) at the onset followed immediately by the Rayleigh-Taylor Instability (RTI) \cite{Richtmyer1960, Meshkov1969, Rayleigh1900, Taylor1950, Sedov1993, Needham2010, Miles2004}. Any initial perturbations present on the interface create local misalignment between the density gradient and the driving, time-dependent pressure gradient, resulting in vorticity deposition on the interface due to the RMI and RTI. The vorticity deposition causes the perturbations to grow into inter-penetrating spikes (high density gas) and bubbles (low density gas). We refer to these combined effects as the Blast-Driven instability (BDI). For a comprehensive and up-to-date survey of research on the RMI and RTI, see \citet{Zhou2017a, Zhou2017b}.

Combined instabilities such as the BDI have been shown to be particularly troublesome for ICF; making their elucidation a crucial step for the viability of this technology \cite{Rana2017, Goncharov2000, Miles2004a, Hurricane2014, Meezan2017, Ma2013}. Laser drive asymmetry and fabrication irregularities create perturbations that grow due to combined instability during the initial implosion stage and later on when compressed hot fuel decelerates the colder imploding fuel \cite{Ribeyre2004, Hurricane2014, Meezan2017, Ma2013}. While RM growth within the capsule has been shown to have a small effect, it serves to initialize the dominant RTI growth of the implosion phase \cite{Zhou2019, Landen2016}. The instability we study is analogous: RMI acts immediately to set the stage for the longer time RTI growth. 

Furthermore, it has been found that the ICF operating regime is located on a performance cliff, meaning small deviations in model design could have a large impact on performance \cite{Rana2017}. This predicament provides impetus to rigorously test the simulations, and their models, used in ICF design as much as possible. From a more general standpoint, the simulation of compressible and turbulent environment such as the BDI pose large challenges for simulations. This is primarily due to the fact that the numerical techniques used to handle shock discontinuities often impact the method's ability to resolve turbulence \cite{Johnsen2009}.

\subsection{\label{subsec:novelty}Novelty of Present Work}

The experimental data on their own provide three unique opportunities for model validation. Primarily, RTI and RMI have been studied extensively in the past decades, but most often in planar geometries. The models, theories, and key governing parameters (such as the self-similar growth constants), were mostly developed with planar studies. Because these models and theories have often times not been developed or validated in cylindrical or spherical geometries, they can introduce errors when used in those regimes \cite{Rana2017}. The present work addresses this by providing experimental data in cylindrical geometry and a more complex test case for simulations, RMI/RTI models, and theories. Second, the experiment provides a combined RMI/RTI instability with which to validate models. The non-linear coupling of instabilities have previously been difficult to predict \cite{Olson2011}. Lastly, the experimental data was taken using high speed (4.75-7 kHz) methods and is temporally well resolved. This allows for high-fidelity simulation validation in that the entire temporal evolution of the instability can be compared directly from experiment.

Beyond the novelty of the experimental data themselves, the simulations in this study represent a high-fidelity representation of the experimental configuration. With both simulations and experiment completed by the authors, nuances of the experimental initial conditions, boundary conditions, and the data acquisition and processing were fully replicated in the simulation. An important example of this is the characterization of interface initial conditions (ICs). Many simulations do not use real ICs to seed their simulations, while those that do often use an idealized version of an experimental IC \cite{Mueschke2009}. For example, in \citet{Boureima2018} a cosine superposition with an idealized Gaussian distribution of initial amplitudes and phases was used to seed multi-modal perturbations. The use of poorly characterized ICs is particularity problematic for transitioning instabilities like the BDI, as several works have shown the imprint of ICs lasting into the late time flow properties and significantly impacting the flow evolution \cite{Mohaghar2017, Thornber2010, George2004, Dimonte1999, Balakumar2012, Zhou2019, Mueschke2009, Zhou2017b, Kadau2007, Ramaprabhu2005, Banerjee2009, Youngs2013}. The ICs in this study were characterized using two different methods. One method used the energy spectra (from $\approx$ 300 IC realizations), and the other used direct data input of the experimental ICs. This process is covered in more detail in Section \ref{sec:les}.

The present study offers novel model comparison and validation through ensembles of data. More specifically, the experimental data was used to produce mean and variance profiles for the quantities of interest. These mean profiles were used to tune the Euler and Reynolds-Averaged Navier-Stokes (RANS) stages of the model validation. However for the Large Eddy Simulation (LES), we use the same number of LES realizations as there are experimental realizations. Then, both the mean and the variance of the LES ensemble are compared to that of the experiment. It is in this way that the study attempts to more accurately recreate the high sensitivity to ICs in this non-linear instability.

To the authors knowledge, no previous works have used this level of ensemble based comparison between experiment and multiple turbulent models. For instance, \citet{Barmparousis2017} compared ensembles of RANS and LES models to study RMI uncertainty, but did not involve experimental observations or ICs. \citet{Clark2003} and \citet{Zhou2013} both simulated an ensemble of synthetic/artificial ICs using LES, but the studied RT flows had no link to experimental observations or RANS models. \citet{Narayanan2018} compared simulated RMI to experimental shock tube data, but neither the experimental data or simulation data were ensembles. Finally, over 20 years ago \citet{Dalziel1999} performed an ensemble of RTI experiments and compared them to LES simulations from two different ICs, but the comparisons were primarily   qualitative and the ensembles themselves only consisted of a few runs. Indeed, it seems extremely rare to find examples of ensemble studies for instability and mixing problems, let alone those also linked to experimental data.

Finally, the model validation occurred in an hierarchical approach. First, the initial conditions and boundary conditions are set by matching 0th order experimental data (such as pressure profiles) to results from the direct solution of the Euler equations, i.e the Euler model. The Euler simulation tuning parameters are then set as constants for the mix-model simulations. A $K-L$ RANS model \cite{Morgan2022} is validated by tuning three model parameters to best match experimental data like interface mixed-width. The second mix-model investigated is a 3D LES model, which is validated using an ensemble approach as mentioned above. All three models used in this study are implemented using the Pyranda code, an open source proxy-app for the Miranda code developed at Lawrence Livermore National Lab.

In summary, the goal of the present work is to create and demonstrate a high-fidelity numerical model of our experimental facility in order to test and validate RANS and LES models through comparisons of the ensemble data.  The fidelity of the numerical model will be high enough for it to act as a "digital twin" and be used for future experimental design work. This could in turn improve the confidence in results when using these models in much less controlled problems, such as ICF or SN. The paper is presented in the following manner. First the experimental facility and data are introduced and explained. Next the Pyranda code and the equations of motion are introduced, along with the staged-validation approach. The subsequent three sections cover the specific validation steps for each of the three models used (2D Euler, 2D RANS and 3D LES) with their corresponding experimental data. The final section discusses the results of the validated simulation.

\section{\label{sec:facility}Experimental Facility}

\subsection{\label{sec:data}Facility and Data Overview}

The experimental facility used to acquire the validation data has been detailed and characterized previously in \citet{Musci2020}, but will be briefly summarized here. Figure \ref{fig:facility} shows a photo, schematics, and examples of initial conditions produced by the experimental facility. As shown in Figure \ref{fig:facility}a the chamber is triangular in shape as it is intended to emulate a sector of a cylindrical disk. It consists of two 45$^{\circ}$ diverging steel plates with about 3 cm of spacing in-between. The experiments take place in the center of this space. The chamber is over 2 m tall to allow for longer experimental times by minimizing reflected rarefaction waves from the top, which is open to the ambient. 

Figure \ref{fig:facility}b-c show schematics that explain the operation of the facility. Two gases of different density fill the chamber - the high density gas ($\rho_{Heavy}$ or $\rho_{H}$) enters from the bottom, while the low density gas ($\rho_{Light}$ or $\rho_{L}$) enters from the top. The gases exit 0.5 m above the chamber bottom through a horizontal slot due to suction created by a fan attached to the outer wall of the chamber. The fan causes slight variations in the initial gas velocity which introduces small perturbations, $\mathcal{O}$(mm) amplitudes, along the interface. The facility can also be operated such that the interface remains flat and unperturbed, as is show in Figure \ref{fig:facility}d.i. Otherwise, Figure  \ref{fig:facility}d.ii-vi shows examples of perturbed initial conditions. The perturbed interface can exhibit a range of different amplitudes and wavelengths from run to run but nominally produces perturbations with initial amplitudes ranging from 1-8 mm and wavelengths between 2-6 cm.

The facility generates blast waves using a commercial detonator (RP80 from Teledyne RISI) placed at the bottom vertex of the chamber. Upon detonation a BW travels upward, impacts the interface, and causes instantaneous hydrodynamic development. The BW has a Mach number of approximately 1.6 before interface impact. The large width to depth ratio of the facility (62 cm wide at the interface location gives a ratio of 21) helps to reduce any out-of-plane dynamics and keeps the dynamics approximately 2D (except at the fine scales, where 3D turbulence develops). Additionally, in order to isolate the physics of interest, a thin layer of foam is added along the facility side walls, as indicated in Figure \ref{fig:facility}b. This has been observed to eliminate the reflected bow shocks resulting from the BW-side wall interaction without impacting the hydrodynamic behavior at the center of the facility. 

For all flow visualization diagnostics, the high-speed cameras used are only able to record the physics occurring in the facility's viewing window. The relative size and location of this window is indicated in Figure  \ref{fig:facility}b-c and can also be seen in Figure \ref{fig:facility}a. For the present study three types of data were used. 1.) Dynamic pressure transducers/probes (DPT - model PCB 113B27) were used to record time resolved pressure traces at four locations along the length of the facility. The position of these four probes are indicated in Figure \ref{fig:facility}b and allow for the tracking of the BW front. 2.) High-speed Mie Scattering was used to illuminate the fog seeded heavy gas and obtain large-scale mixing or interface trajectory data within the viewing window. 3.) High-speed Particle Image Velocimetry (PIV) was used to obtain 2D velocity field data for the evolution of the instability. However, in order to achieve adequate resolution, the PIV data was taken from a smaller field of view, located within the main/larger viewing window used for 2.).

\subsection{\label{subsec:modeling}Modeling Challenges} 

\subsubsection{\label{sec:bcs}Boundary and Initial Conditions}

Along with matching the general facility geometry and scale, there were several unique details about the facility which had to be captured by the simulation. First was the method used to simulate the explosion of the detonator at the bottom of the chamber. While the direct simulation of the detonator could be the subject of its own research effort, the goal of this work was to simply induce the generation of a BW with characteristics that matched the experimental data. This was done in the digital twin using a small ``energy pill." A user specified quantity of total energy is placed inside a small sphere which declines steeply (following a radial Gaussian profile) to the ambient at the radius. The motivation behind this method being that releasing a large amount of energy in such a small space, would allow the simulation to form a Taylor-Sedov BW \cite{Grun1991, MacFarlane1989}, which has been shown to be produced by the experiments \cite{Musci2020}. The radius and location of the energy pill were motivated by the actual size and location of the detonator used in the experiments. 

\begin{figure*}[ht!]
\includegraphics[scale = 0.65]{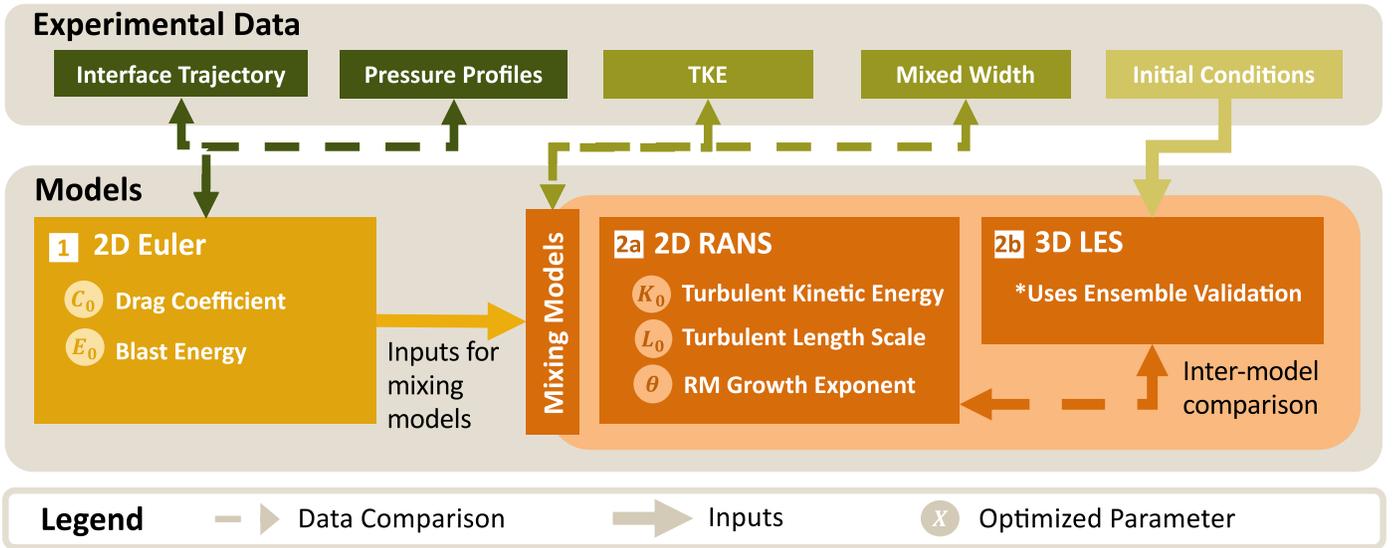}
\caption{\label{fig:fchart} A diagram outlining the staged validation approach taken in this study. \textbf{1)} Stage 1 is considered the baseline solver: the 2D Euler model. The Drag Coefficient, $C_o$ and initial Blast Energy, $E_o$, are tuned in order to optimize agreement with experimental data. The data used for comparison in Stage 1 are the Interface Trajectory and Pressure Profile data. Once optimized, $C_o$ and $E_o$ are set as constants and used as inputs for Stage 2. \textbf{2)} Stage 2 involves validation using Mixing Models. \textbf{2a)} Stage 2a uses a 2D RANS model where the initial Turbulent Kinetic Energy, $K_o$, initial Turbulent Length Scale, $L_o$, and self-similar RM Growth Exponent, $\theta$ are used as tuning parameters to optimize agreement with experiment data. The experimental data used for comparison in Stage 2 are TKE and interface mixed width. \textbf{2b)} Stage 2b involves validation with a 3D LES model. No optimized parameters are used in Stage 2b because the 3D LES model is not tunable. Further, a data-driven  representation of the experimental interface Initial Conditions are used as inputs to Stage 2b. Representative ICs for each experimental realization allow for the 3D LES model to be ensemble validated by comparing the mean and variance data against the experimental data. Finally, the model outputs of Stages 2a and 2b are compared against each other using quantities of interest not available in the experimental data set.}
\end{figure*}

The second modeling challenge is the presence of foam material along the side walls of the experimental facility. The foam which is sandwiched between the two larger face plates, is placed on the diverging side walls and extends roughly two inches into the chamber. This is present to eliminate the reflected shocks that are created as the BW propagates up the chamber. It was seen experimentally that the reflected shocks impacted the hydrodynamics of the instability substantially. Thus, in order to isolate the effects of the incident BW-interface interaction, the foam was added and shown to eliminate all reflected waves. This was emulated in the simulation by placing a custom boundary condition along the edges of the 2D domain. It should be noted that trying to accurately model the physics of the shock-foam interaction was not the goal; it was simply desired to eliminate reflections off the side wall. This was accomplished by applying a low pass Gaussian filter (as in \citet{Cook2007}) to the flow field in the same physical location as the foam in the experiment.  The velocities at the walls were set to zero, and transition between the filtered foam region and unfiltered flow region was defined using a hyperbolic tangent function with a thickness of a few computational grid points.  

Similarly, the outflow BC had to be modified to match that of the experiment. The experimental facility has an open lid, such that the BW passes through the chamber and freely into the ambient. The chamber height was made such that by the time the BW exits the chamber it has weakened substantially. The weakened BW exits the facility and sends a reflected rarefaction wave back into the chamber. The rarefaction has been observed to impact the hydrodynamics of the instability, minimally, at late times ( $>$ 11 ms). To mimic this feature in the digital twin, the outflow was formed so that minimal reflections would be created at the top boundary. This is achieved by filtering the flow as it approaches the top of the domain and enforcing ambient conditions at the boundary.

\subsubsection{\label{subsec:losses}Losses}
Unmeasured losses in the experimental facility undoubtedly affect flow field physics, and so must be accounted for in the digital twin. For instance, prior to accounting for losses the simulations displayed excessive interface movement as compared to experimental observations. The extra, late-time, movement of the interface indicated that more energy losses were occurring in reality than was being produced by simulations. 

The energy losses are partially due to material/plate deflection both when the detonator explodes and as the BW traverses the chamber. Outflow losses could also occur along the length of the chamber as the BW passes things like modular port plates or viewing windows, and the high pressure forces fluid out of imperfect seals. However, a simple deflection analysis showed that the losses from the pressurized fluid interacting with the plates was likely small compared to those due to turbulent boundary layers. 

Thus, energy losses due to the formation of turbulent boundary layers along the faces of the large 45$^{\circ}$ plates were assumed to be the primary source of loss and motivated the loss-model. The boundary layers have not been experimentally measured but certainly dissipate some flow momentum into internal energy. The combination of losses acting in the system are non-linear and cannot be easily accounted for by simply changing the input energy delivered by the detonator. To account for this in the digital twin, a drag model was used to create a loss term in the equations of motion, which then accounted for the excess energy that was being delivered to the interface in the model (which was not there in the experiment). This model will be discussed further in Section \ref{subsec:gov_eqns}. Those wishing to skip ahead to the results and validation section should see Section \ref{sec:euler}.

\section{Model Details and Methods} \label{sec:model}

The simulation tool used in this work is Pyranda - the open-source proxy app for the Miranda code developed at LLNL. Pyranda uses the same 10th-order central finite-difference scheme in space and explicit 4th order Runge-Kutta time integration scheme as those used in Miranda. Full details of the numerical method used in Miranda are available in \textcite{Cook2007}.  Pyranda (\url{https://github.com/LLNL/pyranda}) is a Python based, hyperbolic PDE solver capable of solving the 3D compressible Naiver-Stoke equations. For numerical discontinuities (shocks, contacts, material interfaces) we use an artificial diffusivity method similar to that of \textcite{Cook2007} to regularize the solutions.  The exact form of these diffusivities is given in \textcite{Morgan2018c} and in Section \ref{subsec:gov_eqns} below. 


All simulations are run on a 2D (Euler and RANS) or 3D (LES) conformal structured mesh, where grid spacing is much finer near the bottom vertex of the chamber domain and gets progressively coarse in the vertical direction. The mesh is designed such that there are twice as many grid points in the vertical direction as there are in the horizontal.

\subsection{Staged Validation Approach} \label{subsec:approach}

An aspect of novelty in the current study is the approach taken for model validation. As opposed to validating a single RANS or LES mix-model against experimental results, this study validates both simultaneously. Due to this, the model validation occurs in a staged approach, whereby a non-mixing, 2D Euler validation is completed before moving on to the mix-model validation.   

Figure \ref{fig:fchart} shows the staged validation approach as a schematic. The bottom of Figure \ref{fig:fchart} shows a legend and the top shows the four types of experimental data used to assess model performance. The four experimental data types exist in two classes: non-mix and mix data. The non-mix data consists of pressure profiles obtained from the DPTs during \textit{single}-gas runs, and interface trajectory data obtained from high speed Mie Scattering during two-gas runs with a flat IC.  This non-mix data was used first to validate the 2D Euler model, which is itself a non-mixing model. 

The mixing models, on the other hand, are both validated against the experimental mix data, which consists of high speed PIV data in the from of Turbulent Kinetic Energy (TKE or $k$) estimations and integral mixing-width ($W$) data obtained form high-speed Mie Scattering. All experimental mix data was for runs with a perturbed IC. Thus, the four data just outlined are also the Quantities of Interest (QOI) which models will output for comparison. To summarize, the four primary QOI used for model-experiment comparison are: pressure probe data, flat interface trajectory, integral mixed-width ($W$ - Equation \ref{eq:w}), and TKE ($k$- Equation \ref{eq:exp_tke}).

The tuning parameters used in each of the three computational models are shown as circles in the boxes of Figure\ref{fig:fchart}. How these parameters make their way into the model's governing equations are shown in Section \ref{subsec:gov_eqns}.

In the Euler model validation (Stage 1), two model parameters were used to optimize the simulation output against the non-mix experimental data. These parameters are indicated by the circles in the ``2D Euler" block in Figure \ref{fig:fchart}, and are the Drag Coefficient, $C_o$, and the initial Blast Energy input to the energy pill, $E_o$. These parameters were varied in both single-gas and two-gas Euler simulations until they produced an optimum solution when compared to the experimental pressure and interface data. The optimum solution was determined by minimizing the error between the simulation output and the experimental data, and is outlined in Section \ref{sec:euler}.  

In Stage 2, the mixing experiments utilizes the same detonators and divergent geometry as Stage 1, $C_o$ and $E_o$ are set to constants and used as inputs to the ``Mixing Models" shown in orange in Figure \ref{fig:fchart}. Comparison of the RANS and LES models is the next stage (2a and 2b) in the validation and can be performed concurrently. The mixing models are both validated against the experimental mix data, which consists of high speed PIV data in the from of TKE measurements and integral mixing width data obtained again form the high-speed Mie Scattering. All experimental mix data was for 27 different runs, each with a perturbed IC. 

It is important to note that the 2D Euler computational model serves as the base code for both the RANS and LES mix models. The Pyranda framework allows for additional capabilities, degrees of freedom, transport equations and more to be integrated into the base Euler model code.  This allowed us to maintain a single model and source code defining the hierarchy which has proved valuable in ensuring consistency.

As seen in the orange ``RANS" box in Figure \ref{fig:fchart}, the RANS model was optimized using three model parameters: initial turbulent kinetic energy, $K_o$, initial turbulent length scale $L_o$, and the self similar RMI growth exponent, $\theta$. These three parameters were varied to optimize the simulation outputs against the experimental mix data, namely windowed TKE and integral mixed width.

Comparisons of the LES model to the experimental mix data did not involve determining any optimal model parameters like the previous stages, as shown in Figure \ref{fig:fchart}.  The LES model, however, requires that the ICs of the interface be explicitly defined.  Given the strong dependency of the mixing on the ICs, an ensemble of representative samples of the ICs were required in both the experiment and simulation in order to generate significant means in the data.  Two approaches were taken in initializing the LES ICs.  The first used a statistical model of the experimental ICs and  then sampled from that model to generate the ICs for the LES.  The second approach directly replicated the interface profiles of the 27 experimental runs as the LES ICs.  Both approaches showed good agreement of the ensemble mean mixing width versus time.  The direct data approach, however, appeared to be able to match the experimental ensemble variance far better, as will be illustrated in Section \ref{sec:les}.
Lastly, the simultaneity of validation for RANS and LES, allows for inter model comparison between the two. An LES data-set that has compared favorably to the available experimental data can then be used to compare and improve RANS models where experimental data are lacking.  

A final point regarding the simulation data acquisition methods should be made here. In order to most faithfully compare simulation outputs with experimental data, the process of data acquisition in the simulation sought to exactly mimic that in the experiment. For the non-mix pressure profile data, that meant creating simulation ``probes" and placing them at the same locations as the DPTs in the experimental facility. Pressure data from these probe locations were then sampled at a rate exceeding that in the experiment. All other experimental data was captured via cameras acquiring images from the viewing window in the center of the facility. Due to this, an identically sized and located ``window" was created in the simulation domain, and all data used for comparison had to be sampled from that region alone. For instance, Mie scattering data were simulated by generating images from the mass fraction field at the same resolution and domain extent as the experimental photographs.  These images are then post-processed with the same scripts as experimental data. Similar routines were used for the simulation acquisition of interface trajectory data, integral mixed width, and TKE.


\begin{table}
\caption{\label{tab:table2} Table summarizing the different spatial dimensions and terms used in the EOM for each of the validated models.}
\begin{ruledtabular}
\begin{tabular}{ccccccc}
 &Dim. &$\Gamma_{Y_{k,j}}$& $\Gamma_{u_i}$ & $\Gamma_{\tau_{ij}}$
 & $\Gamma_E$ & $\Gamma_{q_i}$ \\[3pt]
\hline \\

Euler & 2 & 0 & Eq. \ref{eq:e_loss} & 0 & Eq. \ref{eq:e_loss} & 0 \\[2pt]

RANS & 2 & Eq. \ref{eq:r_gam_y} & Eq. \ref{eq:e_loss} & Eq. \ref{eq:r_gam_t} & Eq. \ref{eq:r_gam_e} & Eq. \ref{eq:r_gam_q} \\[2pt]

LES & 3 & 0 & Eq. \ref{eq:e_loss} & 0 & Eq. \ref{eq:e_loss} & 0 \\
\end{tabular}
\end{ruledtabular}
\end{table}

\subsection{Governing Equations of Motion} \label{subsec:gov_eqns}

All equations used in each of the three models are included below. The implementation of each model necessitates changes to the base equations introduced in this section. These changes, such as dimensionality or term values, will be specified in the section pertaining to that model. Of particular note are the $\Gamma$ terms on the right hand side of most equations. These terms indicate source (or sink) terms which change values depending on the model implemented and will be discussed in the respective section.  

The general governing equations for the mass, momentum and energy of this compressible, non-reacting, mulit-component model are as follows.

\begin{equation}
\frac{\partial}{\partial t}\left({\rho Y_k}\right) = - \frac{\partial}{\partial x_i}\left({\rho Y_k u_i \: + \: J_{k,i}}\right) 
\label{eq:e_mass}
\end{equation}

 
\begin{equation}
\frac{\partial}{\partial t}\left({\rho u_i}\right) = - \frac{\partial}{\partial x_j}\left({\rho u_j u_i - \tau_{ij}}\right) + \rho g_i + \Gamma_{u_i}
 \label{eq:e_mom}
\end{equation}

\begin{equation}
\frac{\partial E}{\partial t} = - \frac{\partial}{\partial x_i}\left({E u_i - \tau_{ij} u_i - q_i }\right) + \rho g_i u_i + \Gamma_E
\label{eq:e_energy}
\end{equation}

Here $\rho$ is density, $Y_k$ is the mass fraction of species $k$, $t$ is time, $u_i$ is the velocity vector, $x_i$ is the spatial vector, and $J_k$ is the diffusive mass flux of species $k$. 
The diffusive mass flux of each species in Equation \ref{eq:e_mass} is given by

\begin{equation} 
\quad J_{k,i} = \rho D \frac{\partial Y_k}{\partial x_i} + \Gamma_{Y_{k,i}}
\label{eq:e_J}
\end{equation}

\noindent where $D$ is the molecular diffusivity, which is approximated as a scalar for species diffusion between two species. $\Gamma_{Y_{k,i}}$ is an added mass flux source/sink.

In Equations \ref{eq:e_mom} - \ref{eq:e_energy}, $\tau_{ij}$ is the viscous stress tensor, $g_j$ is the gravitational body force vector, and $\Gamma_{u_i}$ is an added loss force vector which will be expanded upon below. $E$ is total energy and $\Gamma_E$ is an added energy source/sink term which will also be described below. $q_i$ is the heat flux vector and is described by

\begin{equation} 
q_i = \kappa \frac{\partial T}{\partial x_i} + \Gamma_{q_i}
\label{eq:e_heat}
\end{equation}

\noindent where $\kappa$ is the thermal conductivity, $T$ is the temperature, and $\Gamma_{q_i}$ is an added heat flux term. 

The form of the viscous stress tensor for a compressible fluid can be expanded upon by 

\begin{equation} 
\tau_{ij} = 2 \mu S_{ij} + \left[ \left( \beta - \frac{2}{3} \mu \right) \frac{\partial u_i}{\partial x_i} - p \right] \delta_{ij} + \Gamma_{\tau_{ij}} 
\label{eq:e_tau}
\end{equation}


\noindent where $\mu$ is the dynamic/shear viscosity, $S_{ij}$ is a symmetric tensor representing deformation of a fluid element due to the rate of strain it is subject to, $\delta_{ij}$ is the Kronecker delta, $p$ is the static pressure, and $\beta$ is the bulk viscosity. $\beta$ is normally 0 for incompressible flows and also for compressible flows where the Stokes assumption holds. This assumption being that there is negligible difference between the mechanical and thermodynamic pressures, i.e. where thermodynamic equilibrium is maintained. However, during rapid expansion or compression of a fluid in situations such as the BDI, the molecular distribution of internal energy can be heavily skewed to the transnational modes (as compared to the rotational and vibrational modes) and drive thermodynamic non-equilibrium, where $\beta$ must be greater than 0 \cite{Kundu2015, Panton2013}. The strain rate tensor, $S_{ij}$, can be further expanded to

\begin{equation} 
\quad S_{ij} =  \frac{1}{2} \left( \frac{\partial u_i}{\partial x_j} + \frac{\partial u_j}{\partial x_i}\right) .   
\label{eq:e_S}
\end{equation}






A mixed fluid equation-of-state (EOS) is found using:

\begin{subequations}
\label{eq:e_eos}
\begin{equation} 
\begin{split}
\gamma = \frac{c_{p_k} Y_k}{c_{v_k} Y_k} ,  \quad R = R_u \frac{Y_k}{MW_k} , \\[5pt]  
p = \left( E - \frac{1}{2} \rho u_{i}u_{i} \right) \left( \gamma - 1 \right) 
\end{split}
\end{equation}

\noindent such that

\begin{equation} 
T = \frac{p}{\rho R} .
\end{equation}
\end{subequations}

\noindent Here $\gamma$ is the specific heat ratio, $c_{p_k}$ and $c_{v_k}$ are the constant pressure and volume specific heats of species $k$, respectively, $MW_k$ is the molecular weight for each species, and $R$ is the apparent gas constant of the mixed fluid with $R_u$ as the universal gas constant.

Artificial transport terms are added to the bulk viscosity $\beta$, the dynamic viscosity $\mu$, the thermal conductivity $\kappa$, and the molecular diffusivity $D_k$ for each species $k$. Each artificial term is also Gaussian filtered in the same process as that of  \citet{Morgan2018c} but is also outlined here for clarity. The fluid properties are defined as

\begin{subequations}
\label{eq:artif}
\begin{equation}
\mu = \mu_f + C_{\mu} \;\; \overline{ \: |{\Pi( \; |{ S_{i,j}}| \; )}| \:  } \;\: \rho 
\end{equation}

\begin{equation}
\beta = \beta_f + C_{\beta} \;\; \overline{ \Pi \left(\frac{\partial u_i}{\partial x_i}  \right) \: \rho  }
\end{equation}

\begin{equation}
\kappa = \kappa_f + C_{\kappa} \;\; \overline{  \frac{\rho}{T 
\Delta_t} c_v  \: G(T)  \:  }
\end{equation}

\begin{multline}
D_k = D_{k,f} \; + \\
\overline{  \frac{\rho}{\Delta_t} \; \max \left[ \Pi(Y_k) C_{D_k} , \: \: C_Y \left( |{Y_k}|-1 + |{1 - Y_k}| \right) \Delta^2  \right] }
\end{multline}

\begin{equation}
C_i = 
	\begin{cases}
  		1\times 10^{-3}, & i = \mu \\[3pt]
  		1\times 10^{-1}, & i = \beta \\[3pt]
  		1\times 10^{-3}, & i = \kappa \\[3pt]
  		1\times 10^{-4}, & i = D_k \\[3pt]
  		1\times 10^{2}, & i = Y
	\end{cases}
\end{equation}

\end{subequations}

\noindent where $f$ denotes the physical/real contribution to the fluid transport property and all additional right-hand-side (RHS) terms make up the artificial contribution. For the artificial terms, the overbar indicates the application of a truncated Gaussian filter, $\Delta$ is the local mesh spacing, and $\Delta_t$ is the instantaneous time step of the solver. The function $\Pi$ represents the application of an eighth-derivative operator to the scalar or vector in parenthesis. For some arbitrary scalar, $\phi$, and vector $\phi_i$, $\Pi$ can be respectively expressed as 
 
\begin{subequations}
\label{eq:g}
\begin{equation}
\Pi\left( \phi \right) = \max \left( |{\frac{\partial^8 \phi }{\partial x^8} \Delta x^8}|, \: |{\frac{\partial^8 \phi}{\partial y^8} \Delta y^8}|, \: |{ \frac{\partial^8 \phi}{\partial z^8} \Delta z^8}| \right), 
\end{equation}

\begin{equation}
\Pi\left( \phi_i \right) = \max \left( \Pi(\phi_x), \:  \Pi(\phi_y), \:  \Pi(\phi_z) \right),
\end{equation}
\end{subequations}

\noindent Finally, the artificial transport coefficients are specified in Equation \ref{eq:artif}e. Note that in Equation \ref{eq:artif}a the eighth derivative operator, $\Pi$, is applied to the magnitude of the strain rate tensor, which is equivalent to the square root of the inner product of $S_{ij}S_{ji}$.
    
As discussed in Section \ref{subsec:losses}, sink terms in both the momentum and energy equations (Equations \ref{eq:e_mom} - \ref{eq:e_energy}) were added to account for the unmeasured losses in the experimental facility.  The loss force, $\Gamma_{u_i}$, and energy, $\Gamma_E$, variables were made time and space dependent by tying them to a simple ad-hoc boundary layer evolution equation. This simple model was developed following the canonical laminar and turbulent boundary layer models \cite{Blasius1907,Schlichting2015}. By taking the time derivative of these equations/models and assuming $dx/dt \approx u_{drg}$, the boundary layer width, $\delta$, can be shown to develop in time according to:

\begin{subequations}
\label{eq:e_udrg}
\begin{equation}
\frac{\partial \delta}{\partial t} =
	\begin{cases}
  		5.0 \; |{u_{drg}}| {Re}^{-0.5} , & Re \leq Re_c \\[8pt]
  		0.37 \; |{u_{drg}}| {Re}^{-0.2}, & Re > Re_c 
	\end{cases}
\end{equation}

\noindent where 
\begin{equation} 
Re = \frac{|{u_{drg}}| \; \delta}{\nu} , \quad u_{drg} = \hat{r}_i u_i . 
\end{equation}
\end{subequations}

\noindent Here $\hat{r}$ is a radial unit vector to the origin of the domain, which ensures that only the the portion of the velocity vector, $u_i$, acting along radial lines (i.e $u_{drg}$) contributes to the loss terms. Meanwhile, $Re$ is the Reynolds number and $\nu$ is the kinematic viscosity. 
Note the $Re$ above is not that relating to the instability development, but is solely related to the boundary layer along the chamber walls.

Then, assuming a drag coefficient ($C_{drg}$) should increase proportionally to the ratio of the boundary layer height (on each plate) to the chamber half-width such that $C_{drg} = 2 \, C_o \, \delta / W$. We use this to approximate a drag, or loss, force through:

\begin{equation} 
F_L = 2 C_o \frac{ u_{drg} |{u_{drg}}| \rho \delta  }{W^2}  
\label{eq:e_drgfroce}
\end{equation}

\noindent such that

\begin{equation} 
\Gamma_{u_i} = -\hat{r}_i F_L  ,  \quad \Gamma_E = \Gamma_{u_i} u_i \, . 
\label{eq:e_loss}
\end{equation}
 
\noindent Here, $F_L$ is the per unit volume drag/loss force, $W$ is the experimental chamber half width, and  $C_o$ is a drag/loss coefficient. $C_o$ is one of two tunable parameters used to optimize the loss terms and validate the Euler model against experimental results. It  is discussed further in Section \ref{sec:euler}.

\subsection{2D Euler Model} \label{subsec:euler}

Equations \ref{eq:e_mass} - \ref{eq:e_loss} make up the foundational equations applied to each of the three models used in this study. Stage 1 in this study's model validation is the base level simulation in Pyranda and involves solving of the 2D Euler equations. In the Euler model used in Stage 1 of the validation scheme, all of the foundational equations are applied in 2D using one or two different gas species. Additionally, the following source terms are set to 0: $\Gamma_{Y_{k,i}} , \Gamma_{q_i} , \Gamma_{\tau_{ij}}$.

\subsection{2D RANS Model} \label{subsec:rans_model}

The RANS model used in the Stage 2a validation step was implemented through the use of LLNL's RANSBox package of turbulence models \cite{Morgan2022}. RANSBox couples with the Pyranda source code to calculate the turbulent diffusivities and source terms in the turbulent transport equations. RANSbox adds these turbulent diffusivities, source terms, and transport equations to the foundational equations (Eqs. \ref{eq:e_mass}-\ref{eq:e_loss}) used in the 2D Euler model. The particular RANS model used for this study was a 2D $K-L$ model, which adds turbulent transport equations for a second order velocity moment, $K$, the turbulent kinetic energy, and the turbulent length scale, $L$. This $K-L$ model is derived from the RANS equations for a compressible, non-reacting, two-component fluid-mixture and was first implemented by  \textcite{Dimonte2006}.

While analyzing the results of this study, it should be kept in mind that RANS models assume fully developed turbulent flow through their derivation of the turbulent kinetic dissipation rate equation. As such, they are poorly suited to simulate transitioning flows such as the BDI since they assume an established inertial range and fully developed turbulent flow at all times \cite{Zhou2019, Morgan2018a, Morgan2018c}. Therefore, for transitioning flows such as this, one can only expect RANS to match experimental data well at late times when the flow is more fully developed.

The equations governing the RANS model are the previous Equations \ref{eq:e_mass} - \ref{eq:e_loss} but with  different values for all source, $\Gamma$, terms (Equations \ref{eq:r_gam_y}-\ref{eq:r_gam_e} ). Furthermore, as this is a \textit{Reynolds Averaged} model, all variables represented in Equations \ref{eq:e_mass} - \ref{eq:e_loss} are replaced with their Reynolds, or Favre, averaged counterpart. The Favre and Reynolds average are defined in Equation \ref{eq:r_favre}, but the replaced variables are namely: $u_i \rightarrow \tilde{u}_i, \: \rho \rightarrow \overline{\rho}, \: p \rightarrow \overline{p}, \: Y_k \rightarrow \tilde{Y}_k, \; \text{and} \; E \rightarrow \tilde{E}$. Additional equations for transport and closure of the turbulent variables $L$ and $K$ are also added (Equations \ref{eq:r_k} - \ref{eq:r_closure2}). The added equations and $\Gamma$ terms are summarized below.

The added mass flux term is now expressed as the dissipation of mass flux driven by turbulent species diffusivity. 

\begin{equation} 
\Gamma_{Y_{k,i}} = {\frac{\mu_t}{N_Y}}\frac{\partial \widetilde{Y}_k}{\partial x_i}
\label{eq:r_gam_y}
\end{equation} 

\noindent where $\mu_t$ is the eddy viscosity and $N_Y$ is the species mass diffusivity coefficient. Note that in this $K-L$ model the turbulent diffusivities are constant multiples of the eddy diffusivity, e.g. $\mu_t/N_Y$ in the equation above. This will be true for all turbulent dissipation terms. The $\Gamma$ term in the viscous stress tensor is now



\begin{equation} 
\Gamma_{\tau_{ij}} =  \left[ C_{dev}\left(2\mu_t\overline{S}_{ij}\right)-C_{iso}\overline{\rho}K\delta_{ij} \right] M.
\label{eq:r_gam_t}
\end{equation}

The term in brackets is better known as the Reynolds stress tensor, $R_{ij} = \overline{\rho}\tau_{ij}$, $C_{iso}$ is the coefficient on the isotropic Reynolds stress, and $C_{dev}$ is the coefficient on the deviatoric Reynolds stress.

\begin{equation} 
M = 4 \: Y_k \: \left( 1- Y_k \right)
\label{eq:M}
\end{equation}

\noindent is a measure of fluid mixing in the system and varies from 0-1 in the interface region, ensuring the application of $R_{ij}$ only in the region of the interface. The turbulent dissipation of energy is added through the $\Gamma_{q_i}$ term and is defined by



\begin{equation} 
\Gamma_{q_i} = M{\frac{\mu_t}{N_e}} \frac{\partial  \widetilde{e}}{\partial x_i}
\label{eq:r_gam_q}
\end{equation}

\noindent where $N_e$ is the internal energy diffusivity coefficient and $\widetilde{e}$ is the Favre averaged internal energy, which is $\overline{p} \: \overline{\rho}/(\gamma -1) $.

The $\Gamma$ terms to model the losses in the facility (Equation \ref{eq:e_loss}) are still present in the 2D RANS model. $\Gamma_{u_i}$ remains the same while the turbulent energy source term is added to $\Gamma_E$ such that  

\begin{equation} 
\Gamma_{E} = \Gamma_{u_i} \widetilde{u}_i + M \: \left( C_D\frac{\overline{\rho}\left(2K\right)^{3/2}}{L}  -a_j\frac{\partial \overline{p}}{\partial x_j} \right).
\label{eq:r_gam_e}
\end{equation}

\noindent The term in parenthesis represent the turbulent source contribution to the internal energy. Here $C_D$ is the turbulent kinetic energy dissipation coefficient, $K$ is the turbulence kinetic energy, $L$ is the turbulence length scale, and $a_i$ is the turbulent mass-flux velocity vector.
 
%

The two equations added to the Pyranda equations of motion for this 2D RANS model are the transport equations for $K$ and $L$. These are governed by

\begin{multline}
\frac{\partial}{\partial t}\left(\overline{\rho} K \right) = - \frac{\partial}{\partial x_i}\left({ \overline{\rho} K \widetilde{u}_i + M \frac{\mu_t}{N_k}\frac{\partial K}{\partial x_i}  } \right) + \\[3pt]
{R_{ij}\frac{\partial \widetilde{u}_i}{\partial x_j} \; + \; a_i\frac{\partial \overline{p}}{\partial x_i} \; - \; C_D\frac{\overline{\rho}\left(2K\right)^{3/2}}{L}} 
\label{eq:r_k}
\end{multline} 

\begin{multline} 
\frac{\partial}{\partial t}\left( \overline{\rho} L \right) = - \frac{\partial}{\partial x_i}\left({ \overline{\rho} L \widetilde{u}_i + M \frac{\mu_t}{N_L}\frac{\partial L}{\partial x_i} } \right) + \\[5pt]
C_{L1}\overline{\rho}\sqrt{2K} \; + \; C_{L2}\overline{\rho}L\frac{\partial \widetilde{u_i}}{\partial x_i} \; + C_{L3} R_{ij} \frac{L}{K}\frac{\partial\widetilde{u}_i}{\partial x_j}
\label{eq:r_L}
\end{multline}

The terms on the second lines of Equations \ref{eq:r_k} -\ref{eq:r_L} make up the source terms for the governing equation of each RANS variable, while the terms on the first line containing $\mu_t$ constitute the convective and dissipation terms. Here $N_k$ is the turbulent kinetic energy diffusivity coefficient and $N_L$ is the turbulent length scale diffusivity coefficient. $C_{L1}$ is the coefficient of $L$ production, $C_{L2}$ is the coefficient of $L$ dilatation, and $C_{L3}$ is the coefficient of $L$ shear. 

The remaining RANS terms needed to completely close the equations of motion are shown below: 



\begin{subequations}
\label{eq:r_closure2}
\begin{equation} 
a_i = -C_B\frac{L\sqrt{2K}}{\overline{\rho}}\frac{\partial \overline{\rho}}{\partial x_i}, 
\end{equation} 

\begin{equation} 
\mu_t = C_\mu \overline{\rho}L\sqrt{2K},
\end{equation}
%

\begin{equation}
\overline{S}_{ij} = \frac{1}{2}\left(\frac{\partial \tilde{u}_i}{\partial x_j} + \frac{\partial \tilde{u}_j}{\partial x_i}\right) - \frac{1}{3}\frac{\partial \tilde{u}_k}{\partial x_k}\delta_{ij}. 
\end{equation}
\end{subequations}

\noindent Here $C_B$ is the turbulent mass-flux velocity coefficient, $C_{\mu}$ is the eddy viscosity coefficient, and $\overline{S}_{ij}$ is the deviatoric turbulent strain rate tensor. 

\begin{table*}
\caption{\label{tab:r_coefs} List of the RANS coefficients and their values set by the model for the optimal RANS solution. }
\begin{ruledtabular}
\begin{tabular}{cccccccccccc}
$C_\mu$ & $C_{iso}$ & $C_{dev}$ & $C_B$ & $C_D$ & $C_{L1}$ & $C_{L2}$ & $C_{L3}$ & $N_Y$ & $N_e$ & $N_k$ & $N_L$\\
\hline
1.00 & 0.67 & 0.16 & 0.11 & 1.24 & 6.11 & 0.33 & 0.00 & 6.37 & 6.37 & 6.37 & 3.18\\
\end{tabular}
\end{ruledtabular}
\end{table*}
 
All coefficients used in the RANS equations (Equations \ref{eq:r_gam_y} - \ref{eq:r_closure2}) are listed in Table \ref{tab:r_coefs} and are automatically set by RANSBox through similarity analysis \cite{Morgan2018a, Morgan2022}. It should be noted that the $C_{L3}$ term was set to zero because of the numerical sensitivity it created at early time, due to small values of $K$. This term, not typically included in the canonical $K-L$ models, corresponds to the shear induced by the Kelvin-Helmholtz Instability (KHI), which should have negligible effects early, but may cause discrepancies at later development times \cite{Morgan2018a}.

It must also be noted that in Equations \ref{eq:r_gam_y} - \ref{eq:r_closure2}, an overbar denotes a Reynolds averaged quantity while a tilde denotes a Favre (mass-weighted) quantity. More specifically, for an arbitrary scalar $b$, these averages decompose as 

\begin{subequations}
\label{eq:r_favre}
\begin{equation}
b  =  \overline{b} + b'  =  \tilde{b} + b '', 
\end{equation} 

\noindent and the Reynolds averaged terms can be related to the Favre average through $\rho$ using

\begin{equation}
\tilde{b}  =  \frac{\overline{b\rho}}{\overline{\rho}}  . 
\end{equation}
\end{subequations}

Finally, as this is a instability driven mixing problem the late time, self-similar mixing layer theory can be used to better inform the RANS model. Which for RMI at low Atwood such as this, much research have shown this to be \cite{Zhou2019, Clark2003, Clark2006, Thornber2017, Dimonte2000c}

\begin{equation}
h\left(t \right)  \approx  h_o  \tau^{\theta}   
\label{eq:r_rm}
\end{equation}

\noindent where $h$ is the width of the mixed-layer, $\tau$ is a non-dimensionalized time, $h_o$ is an initial mixed-width, and $\theta$ is a constant: the approximately universal RM growth exponent.

For RTI flows at low Atwood, $A$,(with constant driving acceleration), the self-similar mixed layer has been shown to grow according to \cite{Sharp1984, Cook2004, Glimm2001, Jacobs2005, Zhou2019, Dimonte2004}

\begin{equation}
h\left(t \right)  = \alpha  A g t^2  , 
\label{eq:r_rt}
\end{equation}

\noindent where $g$ is acceleration of the interface, $t$ is time, and $\alpha$ is the approximately universal RTI growth constant. Both $\alpha$ and $\theta$ are tunable parameters in RANSBox. By changing values of these constants, which are more easily compared with experimental data, RANSBox changes all other RANS coefficients (shown in Table \ref{tab:r_coefs}) to best yield the desired RTI or RMI constant. Thus, they serve as a convenient subset of parameters for changing all RANS coefficients en-masse during the validation stage. In this work only $\theta$ is used as a tuning parameter in the RANS model to achieve agreement with the experimental data. 

\subsection{3D LES Model} \label{subsec:les}

Implementing LES allows for the three-dimensional direct capture of the large and intermediate scales of motion in a transitioning flow, while modeling the smallest unresolved scales of motion. 
The LES model implemented here has been used successfully in many RMI and RTI problems before \cite{Morgan2018c, Cook2004, Cabot2006, Olson2007, Olson2011}. For computational efficiency, the 3D LES domain is restricted to only the mixing region of the divergent blast wave facility.  A separate 2D domain covering the entire geometry is also implemented and the two solutions are coupled every time-step on the 3D domain boundaries. This hybrid approach was compared to a full 3D domain approach and QOI results were effectively identical while the computational resources required were reduced by a factor of 8.

The governing equations for the 3D LES model are Equations \ref{eq:e_mass} - \ref{eq:e_loss}, but expanded in all three spatial dimensions. The loss terms $\Gamma_{u_i}$ and $\Gamma_{E}$ are identical to those in the 2D Euler model and are described by Equation \ref{eq:e_loss}. All other $\Gamma$ terms are set to 0. For the 2D LES model applied outside of the 3D interface region, the solved equations are identical in all ways to those used in the 2D Euler model. 
  
The numerical method used to obtain artificial fluid properties (Equations \ref{eq:artif}-\ref{eq:g}) serves as a sub-grid scale model for shear viscosity which becomes active when turbulent flow is unresolved on the mesh.  Positivity in this term ensures that dissipation of turbulent energy, though unresolved, will mimic the physical mechanism of resolved flow albeit on a larger scale.  The artificial shear viscosity also has the quality of being high-order, and will vanish with 8th order convergence when the flow becomes resolved, leaving only physical viscosity as the relevant diffusivity. The artificial diffusivity constants in Equation \ref{eq:artif}e remain the same.

\section{Euler Validation} \label{sec:euler}

The process of validating the non-mixing 2D Euler was outlined in Section \ref{subsec:approach} but will be detailed further here. As indicated in Figure \ref{fig:fchart}, the QOI used for optimization were BW pressure time histories recorded at 4 locations along the chamber, and the ensemble mean of interface trajectory data for the flat IC case. The model parameters used for optimization were the initial blast energy of the ``energy pill," $E_o$, and the drag/loss coefficient, $C_o$. Both these parameters are essentially controls on the uncertainties that exist in the model energetics. 

\begin{figure*}
\centering
\includegraphics[scale=0.52]{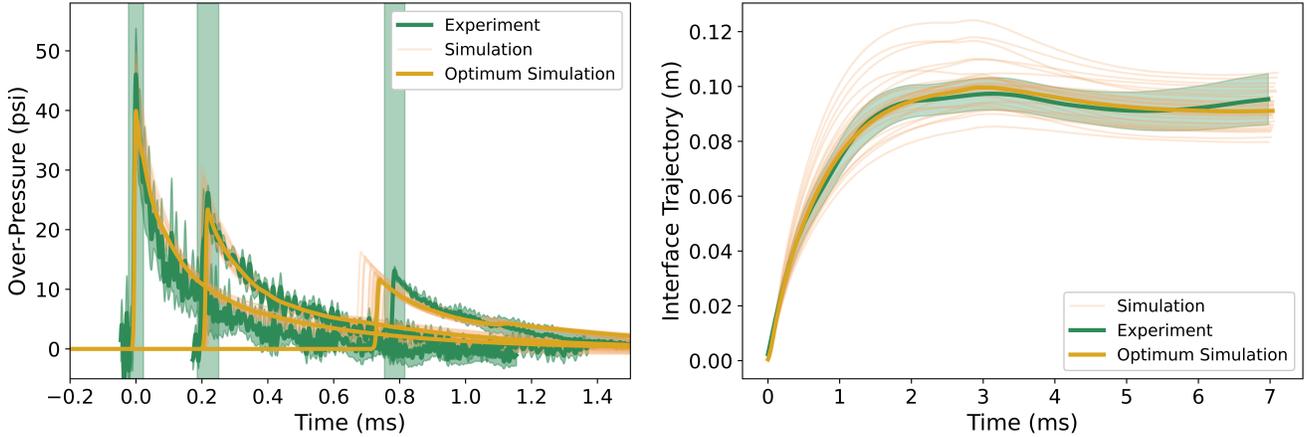}%
\caption{\label{fig:e_pandt} 2D Euler Model: A parameter sweep in $E_o$ and $C_o$ is compared to experimental a) pressure and b) interface data. In both plots the red solid lines indicate the ensemble mean of the experimental data, while the shaded red region indicates the region of plus or minus one standard deviation. The black lines are results from the simulation sweep, with each line indicating a separate $E_o$ - $C_o$ pair. The green line indicate the simulation optimal (minimum error) solution.  }
\end{figure*}

As outlined in Section \ref{subsec:losses} unquantified losses occur during experimental runs. These losses were highlighted by initial attempts to tune the 2D Euler model using only $E_o$, which resulted in interface trajectories that exhibited too much mid-late time movement. This indicated the presence of a loss term that varied in both time and space which could not be accounted for by simply lowering the amount of initial energy in the energy pill. Thus, an ad-hoc drag model was introduced to create a time and space varying loss term, tuned using $C_o$. These losses (which may have included detonation energy lost to the facility's plates and junctions, and energy lost to the developing boundary layer on the large plates), are primarily mix-model independent. As such, they are determined independently in the Euler simulation through $C_o$ and $E_o$. 

The model QOIs used for comparison with the experimental data had to be obtained through two separate sets of simulations. This was because the experimental pressure data existed only for single-gas experimental runs, i.e. the chamber was filled completely with air and the standard detonator was used. Meanwhile, the interface trajectory data were collected during separate runs using two gases, hence the interface. This means that for each set of $C_o$ and $E_o$, a pair of simulations was completed: a single-gas and a two-gas Euler simulation. 

For the single-gas simulations, agreement of the pressure profiles at each probe location was sought. The primary focus, however, was the matching of pressure profiles at the 2nd and 3rd probes, due to the fact that the interface location was between these probes. The pressure probe error between simulations and experiment data were assessed using the time-of-arrival at a probe location, as well as the pressure/amplitude history recorded at each probe. 

The two-gas simulation results were assessed by comparing the flat interface trajectory data. As was already mentioned at the end of Section \ref{subsec:approach}, the simulation data were sampled from an artificial window region in the domain, which matched the physical viewing-window in the experimental facility. In both cases, 2D fog concentration, or mixture-fraction, data was span-wise averaged to create 1D profiles of $Y_H$, the mass fraction of the heavy gas. The 50$\%$ location of these profiles was then recorded at each time to obtain the interface trajectory. The interface trajectory error between simulation and experimental data was estimated using the cumulative absolute difference between the simulation and experimental curves, and then normalizing by the area under the experimental curve. The results of the simulation parameter sweep as compared to the experimental data, are shown in Figure \ref{fig:e_pandt}.

\begin{figure}
\includegraphics[scale=0.47]{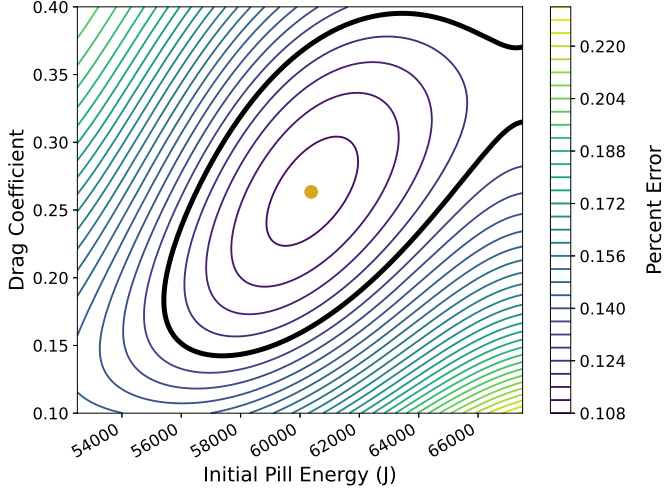}%
\caption{\label{fig:gp_euler} 2D Euler Model: The contour of the GP error surface is shown. The orange dot indicates the realized simulation minimum, which is also the GP predicted minimum in this case. The bold contour line indicates the bounds in which a parameter pair could be selected and still produce results within the bounds of one experimental standard deviation.}
\end{figure}

Simulation results were compared to the experimental data and an estimate of total simulation error was made for each pair of parameters, where the error calculation for each QOI was outlined above. The total error was broken into three contributing factors. First, weighted at 25$\%$ the BW time-of-arrival error was computed using the pressure profiles. Second, and also weighted at 25$\%$ was the error resulting from the differences in peak pressure for each of the pressure probes. Last, and weighted at 50$\%$, was the error for the interface trajectory. A parameter sweep in $C_o$ and $E_o$ was performed using N and M values, respectively.  This resulted in 2*N*M calculations being computed where the relevant QOIs were generated and stored in a database.  Using the total error value for each parameter pair, a Gaussian Processor(GP)  was used to create a high-resolution error surface for all potential parameter pairs ($E_o$, $C_o$). The GP was used to predict, and then test, the optimum pair of parameters which led to the global minimum on the error surface. The results of the GP are shown in Figure \ref{fig:gp_euler}.

As Figure \ref{fig:gp_euler} shows, the parameter pair that was tested to have lowest error, falls nearly exactly where it was predicted to be by the GP. The gold dot in Figure \ref{fig:gp_euler} corresponds to the ``Optimal Solution" lines in Figure \ref{fig:e_pandt}. It should also be noted that the bold contour in Figure \ref{fig:gp_euler} represents the bound on parameter pairs that would result in simulation data falling within one standard deviation of the experimental data. The fact that this encompasses a large array of parameter pairs shows that the simulation results are quite resilient to relatively ad-hoc model choices. The optimal GP solution gave an $E_o$ of 60380J and a $C_o$ of 0.263. These values were then set as constants in the proceeding RANS and LES validation studies.  

\section{RANS Validation} \label{sec:rans_val}

The RANS validation (Stage 2a in Figure \ref{fig:fchart}) takes the optimized parameters from the 2D Euler study, and adds a $K-L$ RANS model to the base simulation to obtain mixing data. The specific details of the RANS model used were outlined in Section \ref{subsec:rans_model}.

The parameters which are varied in this stage of the validation were the initial turbulent kinetic energy $K_o$, the initial turbulent length scale $L_o$, and the self-similar RM growth exponent $\theta$. As opposed to the Euler validation, physical bounds existed for these optimizing parameters. $L_o$ must be based on a length scale that is physically relevant to the experimental problem. This meant $L_o$ ought to only be varied from $O\left(1mm\right)$ - $O\left(10cm\right)$ as this spans the range of minimum observed initial perturbation amplitude to maximum observed wavelength. Meanwhile, $\theta$ has been an extensively studied parameter, so it was bounded by the range of values reported in the literature: 0.21-0.66 \cite{Thornber2017,  Zhou2017b, Weber2012, Thornber2010}.

Optimization of the parameters $L_o$, $K_o$, and $\theta$ occurred in a similar manner as the Euler validation: for a given parameter set, error was calculated for each QOI between the simulation and mean experimental data. The parameter sweep in this case used N, M, P values for $L_o$, $K_o$, and $\theta$, respectively, resulting in N*M*P 2D RANS simulations being run.  Again a GP was used to fit the data and produce a high resolution error surface, now in 4D. Additionally, the QOIs and their respective error measurements were also different.

\begin{figure}[t!]
\centering
\includegraphics[scale=0.61]{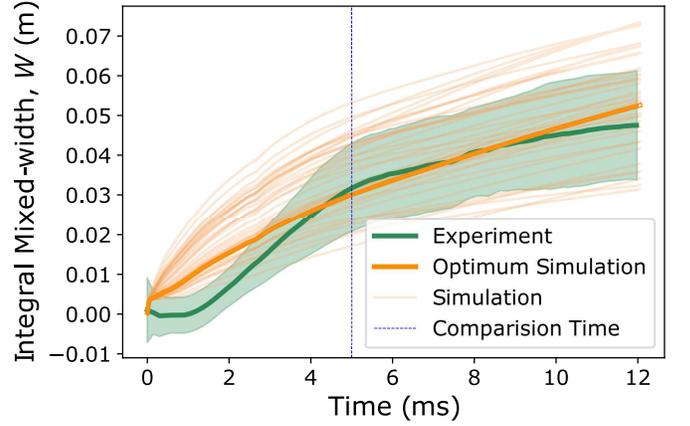}
\caption{\label{fig:rans_mix} 2D RANS model: Mixing data results from a parameter sweep in $K_o$, $L_o$, and $\theta$. The optimized solution predicted by the GP is shown as the bold orange line, experimental data is shown in green and all other parameter sweep results are in light orange. The dashed vertical line indicates the time after which the simulation and experimental data were compared for RANS optimization.}
\end{figure}

The primary QOI used to assess the models ability to capturing the mixing process was the integral mixed width, $W$, found for experiment and simulation following  

\begin{equation} 
W = \int 4 \: \langle Y_H \rangle \langle Y_L \rangle \: dy.
\label{eq:w}
\end{equation}

\noindent where the angled brackets represent averaging in the spanwise (horizontal) direction. By conservation of mass the light gas concentration/mixture-fraction is assumed to be $Y_L =  1 - Y_H$. The 4 is to force $W$ to one at the 50$\%$ mixture-fraction region of the mixing zone. In the experiments, the fog used in the Mie Scattering diagnostic is assumed to trace the heavy gas mixture-fraction $Y_H$. The intensity of pixels in each image, is then directly proportional to $Y_H$. While, strictly speaking, this Mie Scattering technique is qualitative, we use the scattered intensity and the integral value given by Equation \ref{eq:w} as a reasonable proxy to the true mixed width. This assumption is passable in this case because the QOI evaluated is an integrated value. The mixture-fraction field is spanwise averaged at each time, and the resulting 1D profiles are integrated according to Equation \ref{eq:w} to obtain a single value for $W$ at each time step. The time dependent $W(t)$ of each experimental run is then averaged together to obtain the data ensemble used for simulation comparison. An identical process is used to acquire $W$ data in the RANS simulation. The $Y_H$ fields are averaged and integrated at each time to get the $W(t)$ resulting from a combination of simulation parameters, which can then be directly compared to its experimental counterpart. This RANS-experiment comparison is shown in Figure \ref{fig:rans_mix}.

The integral mixed-width error was calculated using the integrated absolute difference between simulation and experiment (just as in the Euler interface trajectory error), but only for a selected time range of the simulation. The first 5 ms after the blast are neglected in the error calculation since the RANS model always assumes a fully turbulent flow where in actuality, the flow is transitional in this region. This explains the early time discrepancies between simulation and experiment in Figure \ref{fig:rans_mix}.  

The second QOI used in the RANS validation was the turbulent kinetic energy. It should be noted that we will generally refer to turbulent kinetic energy as ``TKE." However the TKE values produced by the RANS model will be refereed to as $K$, while the experimental TKE values will be $k$. This distinction is made because these values are closely related but not necessarily precise analogues, as will be discussed here.

The high-speed experimental PIV data, which produces the $k$ data, was acquired from experimental campaigns separate from the Mie Scattering campaigns used for all other experimental QOI in this chapter. As a result, the camera field-of-view was different for the two of these campaigns, and was notably smaller for the PIV/$k$ data. So, a second artificial window was used in the simulation to extract all data that was compared with experimental PIV data. The RANS model's turbulent energy transport variable, $K$, was spanwise averaged in this smaller PIV window, and then the maximum value of each spanwise average was recorded to obtain the temporal development of maximum $K$. 

\begin{figure*}[t!]
\centering
\includegraphics[scale=0.57,center]{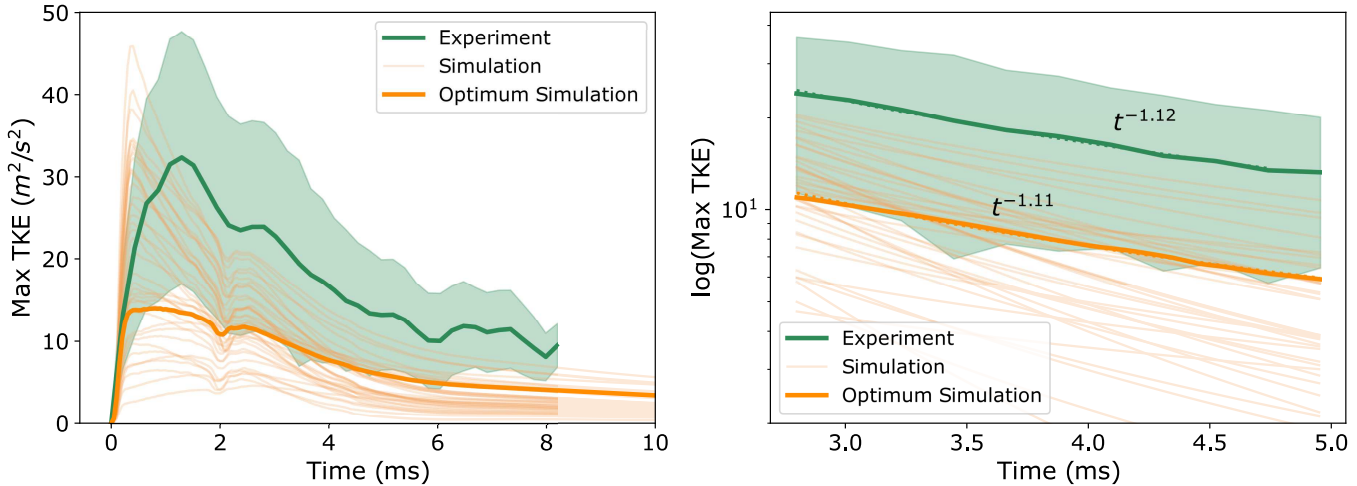} 
\caption{\label{fig:rans_tke} 2D RANS model: TKE results from the parameter sweep in $K_o$, $L_o$, and $\theta$. The optimized solution predicted by the GP is shown as the bold orange line, experimental data is shown in green and all other parameter sweep results are in light orange. a) The left plot shows the maximum TKE of the experiment to be quite a bit larger than the simulation's optimal solution. b) The right plot shows the portion of maximum TKE (2.25-5ms) used to compare the decay exponent ($\Phi$) of maximum TKE. This is plotted in log scale so that lines with identical decay exponents have identical slopes. The optimal solution and experimental slopes agree very well, as indicated by the $\Phi$ values shown on the plot.}
\end{figure*} 

\begin{figure*}
\centering
\includegraphics[scale=0.47]{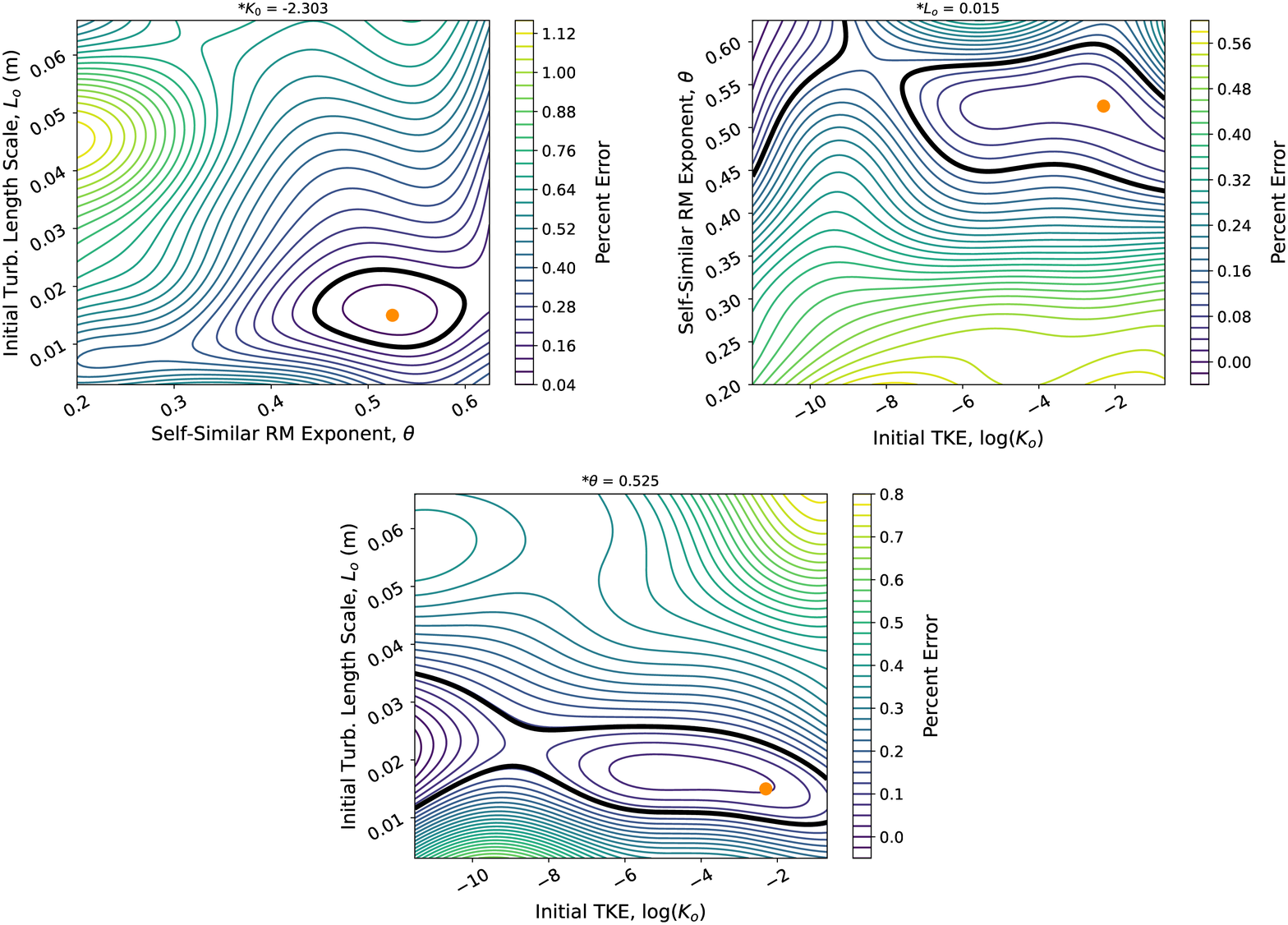}
\caption{\label{fig:rans_gp} 2D RANS Model: The three intersecting planes of the 4D error surface resulting from the RANS GP routine. Note that in each contour plot, two of the optimized parameters are varied, while the fixed/optimal value of the third is indicated at the top of the plot with an asterisk. It should also be noted that the values for $K_o$ are given as $\log(K_o)$ due to the large range in values (10$^{-5}$ - 10$^{-1}$) used in the parameter sweeps. The bold contour line in each plot is once again used to indicate the parameter space which results in simulations within 1 standard deviation of the experimental mean. Clockwise from top left: $\theta$ vs $L_o$, $K_o$ vs $\theta$, and $K_o$ vs $L_o$.}
\end{figure*}

This was compared to the experimental TKE values ($k$), which were computed as follows: 2D fields of velocity vectors were produced using the Davis software from LaVision. The spanwise average components of velocity were then subtracted off the velocity field at each time-step to obtain a field of velocity fluctuations ($v' , \, u'$). These fluctuating quantities were then used to compute a 2D field of $k$ values. Note that due to the assumptions inherent in the RANS model, the $K$ variable it produces must be interpreted as the 3D turbulent kinetic energy, despite the 2D nature of the simulation. Thus to make a comparison to this metric, the 2D experimental data must be modified to create a proxy for the 3D TKE ($K$). This is done by assuming $w' \approx \: u'$, such that:

\begin{equation}\label{eq:exp_tke}
k = 0.5 \sqrt{ 2{u'}^2 + {v'}^2}
\end{equation}

The experimental $k$ field was then spanwise averaged and recorded in the same process as that of the simulation outlined above. However, an additional difference is that the $K$ variable produced by the RANS model assumes a Favre-averaged definition of TKE. In other words, the RANS model outputs a density weighted $K$, which influences the magnitude of the value. In this work, the influence of Favre-averaging on the magnitude of $K$ is assumed to be small. For all of the reasons just outlined, it can only be reasonably expected that, at best, $k \approx K$. These facts are likely a large reason as to why no simulation solutions match the experimental maximum TKE well in Figure \ref{fig:rans_tke}a.

Due to the large discrepancy in TKE magnitudes caused by the differences in $k$ and $K$, the TKE magnitude is not used for the error comparison. Instead, because the energy dissipation rate ($\epsilon$) is determined by the power-law exponent ($\Phi$) of the TKE decay, this exponent is used to compare experiments and simulation. In other words, because $\epsilon \propto t^{\Phi}$, a match between $\Phi_{Experiment}$ and $\Phi_{LES}$ was sought. The decay exponent was only computed during the time range of relatively smooth maximum $k$ decay in the experimental data (from about 2-6 ms). This was selected because it occurs after the small dip in TKE observed in both experiments and simulations. This dip is due to reflected waves, caused by the incident blast-wave passing through the interface, interacting with reflected waves coming from the bottom floor of the facility. The two reflections are initially propagating in opposite directions, but upon collision the resulting acoustic/pressure wave propagates in the direction of the interface. The interaction of this wave with the interface causes the dip in maximum TKE seen in Figure \ref{fig:rans_tke}a at $\approx$ 2 ms. A power law was fit to the data after this interaction, and the resulting exponent of the experiment was used as the target value for RANS optimization and error evaluation.  As can be seen in Figure \ref{fig:rans_tke}b the optimal solution and experimental slopes agree very well, despite the large difference in magnitude. This indicates that the TKE decay exponent in the optimal simulation and the experiment match well, meaning the rate at which they dissipate turbulent energy is nearly the same.

The error from the mixed-width and TKE data were each weighted at 50$\%$ to create the total error that was then more finely sampled using a GP. As mentioned previously, because three optimizing parameters were used ($K_o, L_o, \theta$) the resulting GP error surface is 4D. The minimum error location on the 4D surface was recorded as the optimum solution. The three planes intersecting the error surface at that point are shown in Figure \ref{fig:rans_gp}. Note the values for $K_o$ are given as $\log(K_o)$ due to the large range in values (10$^{-5}$ - 10$^{-1}$) used in the parameter sweep. The lack of variation in $K_o$, however, indicates that the model is relatively insensitive to large changes in the initial turbulent kinetic energy. Additionally, the relatively large array of parameter pairs located within the bold contour in Figure \ref{fig:rans_gp} indicates the RANS model is also quite resilient to model choices.

A final note should be made concerning the optimal RANS solution. The optimized value of $\theta = 0.525$, is relatively high in comparison to the values reported in RMI literature. However, setting this value in the RANS model does not enforce the RMI power law behavior from which $\theta$ gets its name. Indeed, the RANS model detects and responds to finite baroclinic torque, so the RT unstable portion of the BDI will induce appropriate changes (increases) to $\theta$ and mixing rate. Physically, an increased $\theta$ also corresponds to a slower rate of TKE decay as compared to a traditional RMI flow. This is certainly the case in the BDI as the effects of the combined RTI cause further vorticity deposition, and thus more TKE, which can then be sustained longer (i.e. decay slower) due to this second instability.

\section{LES Validation} \label{sec:les}

\begin{figure*}[t!]
\centering
\includegraphics[scale=0.45]{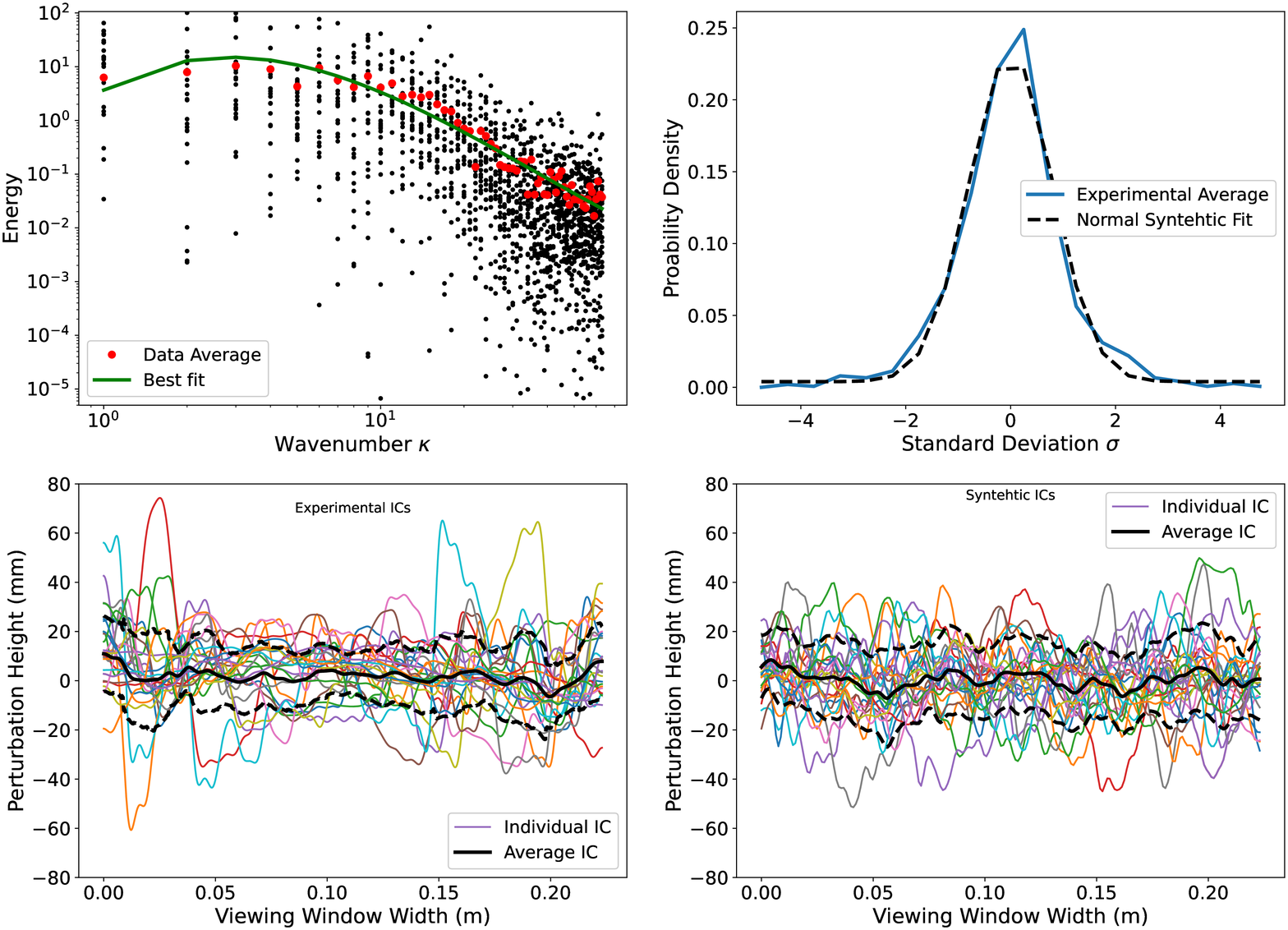}
\caption{\label{fig:les_ic} 3D LES model: Data used in the first IC characterization process. Bottom left - IC shapes from 30 of the 300 images. The mean of the ensemble is shown as the black line, along with one standard deviation. Top left - The result of the DCT for each of the 30 ICs. The red dots show the average energy in each mode, and the line of best fit is shown in green. Top right - The average distribution of energy found at each wavenumber (blue line). The dotted black line shows the distribution fit to the data. The fit distribution is a combined normal and uniform PDF. Bottom right - 30 synthetic ICs resulting from the random sampling of the distribution at each wavenumber.}
\end{figure*}

\begin{figure*}[t!]
\includegraphics[scale=0.6]{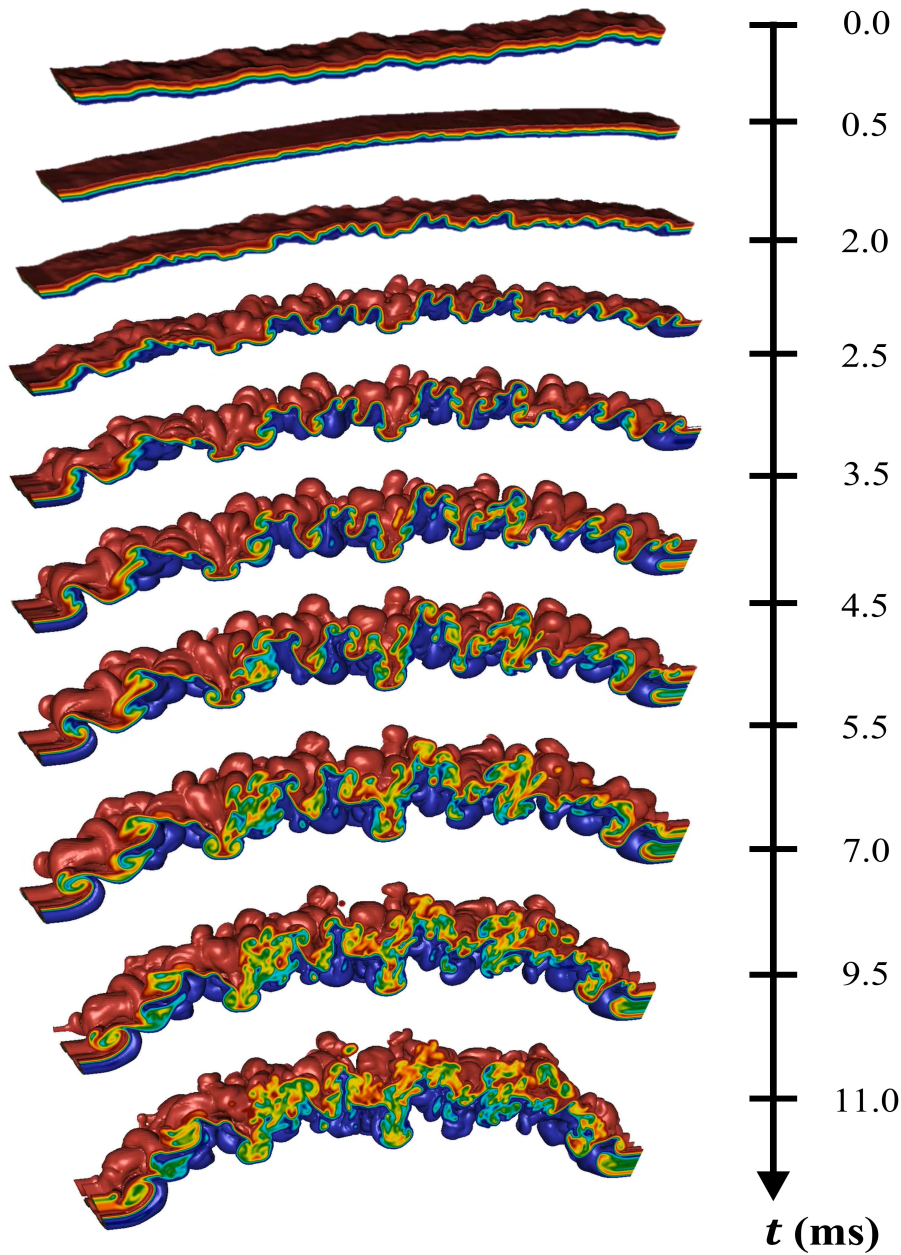}
\caption{\label{fig:les_3d} 3D LES model: A temporal 3D evolution of the $Y_h$ volume fraction field starting from a statistically representative and randomly generated IC. The time axis is shown on the right hand side, with the IC located at the top (t = 0). The blast wave propagates up from below the IC and deposits vorticity to cause the development seen for t > 0. See Supplemental Material at [URL will be inserted by publisher: '3dLES.mp4'] for a video of this 3D LES development. }
\end{figure*}

\begin{figure*}
\includegraphics[scale=0.27]{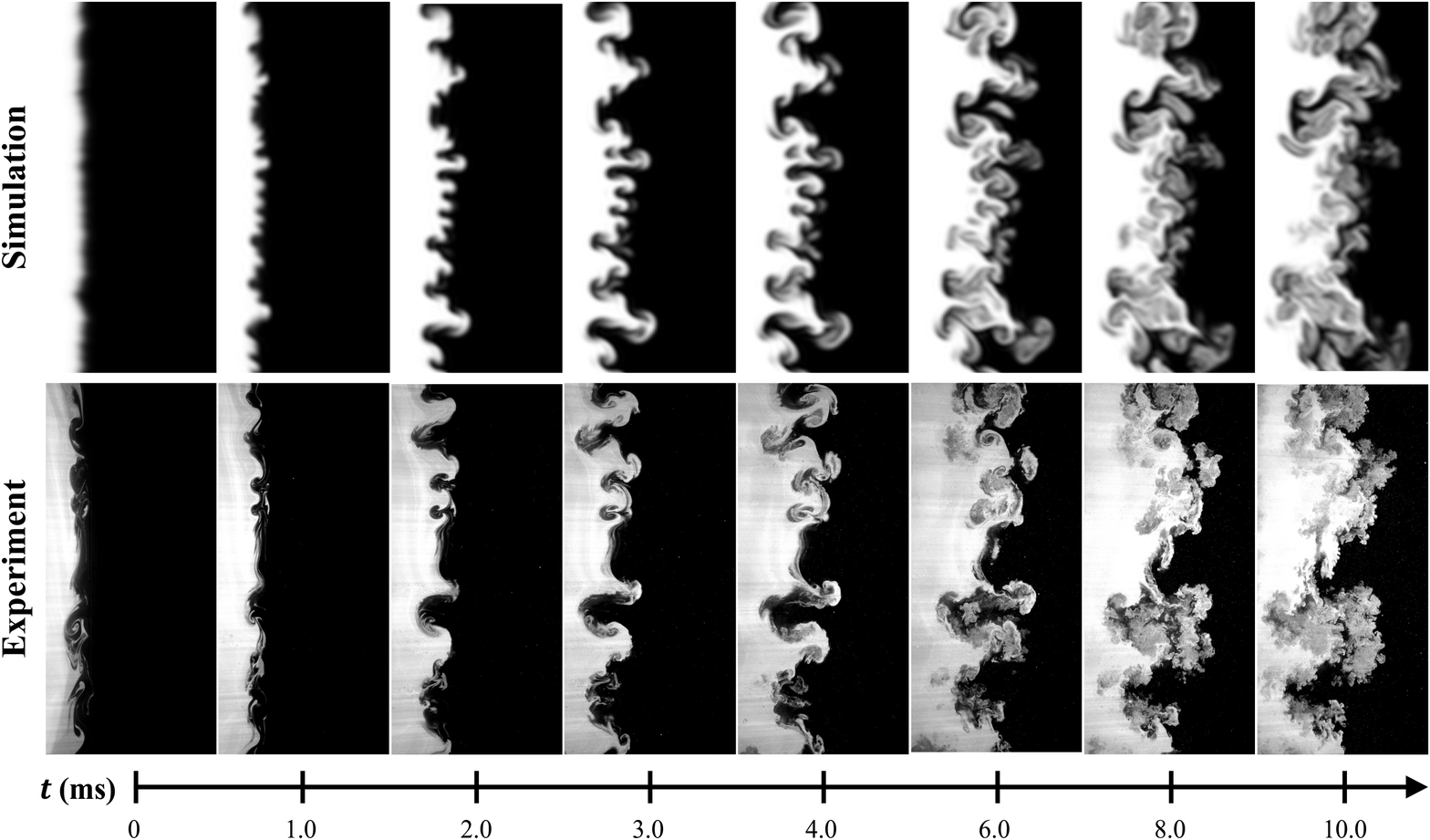}
\vspace{0.01in}
\caption{\label{fig:les_2d} 3D LES model: A comparison of the temporal evolution of the 2D $Y_h$ fields for a slice of the 3D LES and versus the experimental Mie Scattering. Each field is taken from an individual LES or experimental run. The far left Figure is the IC immediately prior to BW interaction. The indicated times (in ms) after blast interaction are identical for LES and experiment. The images have all been rotated 90$^{\circ}$ and cropped to include only the mixing layer at each time step in order to allow for better a better comparison of the instability development. The BW propagates from left to right in the shown orientation and the bulk interface motion, which has been cropped out, is in the same direction. See Supplemental Material at [URL will be inserted by publisher: '2dLES.avi'] and [URL will be inserted by publisher: '2Dexperiment.avi'] for exemplary videos of the 2D LES and experimental Mie Scattering development. Note the two videos are not to scale in size or playback time, and are from different runs than those shown in the images above. }
\end{figure*}

The LES validation (stage 2b in Figure \ref{fig:fchart}) takes the optimized 2D Euler parameters as inputs and solves Equation \ref{eq:e_mass} - \ref{eq:e_loss} but expanded in all three spatial dimensions. The numerical method for artificial fluid properties (Equation \ref{eq:artif}-\ref{eq:g}) serves as an implicit sub-grid scale model for scales smaller than the mesh.

As opposed to the previous validation stages the LES validation requires statistically representative IC information as a model input. This was done using two different approaches. The first method obtains an accurate experimental representation by collecting Mie Scattering snapshots of $\approx$ 300 ``typical" ICs. Parts of the first IC characterization process are shown in Figure \ref{fig:les_ic}, where the plots are shown with data from only 30 of the 300 IC images for better clarity. The first step in the characterization was to obtain an interface shape from each of the IC snapshots by identifying the pixel location corresponding to the maximum gradient of pixel intensity along each column of the image. The resulting 1D array of the interface ``shape," when overlaid on the 2D IC image, traces the interface between the two gases well. Interface shapes from the experimental images are shown in the bottom left of Figure \ref{fig:les_ic}. Next, a Discrete Cosine Transform (DCT) is performed on each of the 300 IC shapes to acquire the energy spectra as a function of wavenumber, $\kappa$, for each experimental IC. The DCT provides information about which ``modes," or $\kappa$, contain the most energy for each IC. The first two modes were removed from each spectra as they corresponded to the mean interface shape (flat), and a ``smile shape" caused by the camera lens' barrel distortion effect at the bottom of each image. A line of best fit is then acquired from this data, as shown in the top left of Figure \ref{fig:les_ic}. The average and standard deviation of the DCT coefficients are then computed for each mode, and are then used to compute a probability density function at each mode/$\kappa$. The distribution created by the IC data was then fit to a combination of a normal and a uniform distribution, an example of which is shown in the top right of Figure \ref{fig:les_ic}. Thus, the line of best fit found for the DCT coefficients essentially sets the mean of each distribution for every resolved wavenumber. In other words, a PDF exists along the Best-fit line in the top left of Figure \ref{fig:les_ic} for each wavenumber.

With a distribution of DCT weights/coefficients at each wavenumber, synthetic ICs can then be generated by cycling through each mode and selecting randomly from the distribution at that wavenumber. With a randomly selected DCT coefficient for each $\kappa$, the inverse DCT can be performed to generate a synthetic IC shape which would be statistically equivalent to a typical experimental IC shape. Resulting synthetic ICs are shown in the bottom right of Figure \ref{fig:les_ic}. However, because 3D ICs are needed for the LES validation, the $\kappa$ dependent randomly selected DCT coefficients are projected to 2D cosine space, and then inverse transformed. This creates a 2D slice of synthetic perturbations that can be overlain on the interface location in the 3D domain. This was done 27 times to generate 27 synthetic ICs in order to match the number of experimental realizations used in the study.

During each of the 27 LES runs, the 3D interface evolves differently due to the random, yet statistically representative, IC generated for each. A 3D evolution for one of the 27 LES runs is shown in Figure \ref{fig:les_3d}. The isocontours shown in the figure represent the volume fraction of the more dense fluid, $Y_H$. The 3D IC is shown at the top of the figure immediately before BW interaction, which travels upward in the figure's frame of reference. In the moments after blast interaction, one can note the apparent compression of the IC initial amplitude. This is caused by the phase inversion, in which initial peaks become valleys (and vice versa), caused by the BW propagating from a more to a less dense gas. Another marked feature is the initial stretching of the overall interface length. The BW forces the interface upwards, and due to the diverging geometry the interface stretches to conserve mass. In addition to the stretching due to geometric effects, the deposition of vorticity due to the subsequent RM and RT instabilities causes local stretching at perturbation sites. By $\approx$ 2 ms the interface amplitude has inverted back to it's initial amplitude, and more than doubles in size in the next 2 ms. This roughly marks the transitions from the exponential growth of the linear phase, to the linear-in-time amplitude growth of the non-linear phase. The beginnings of vortex roll-ups caused by KHI can also be seen at several spike and bubble tips at 3.5 ms. By 7 ms the volume fraction isocontours begin to collide and merge until a turbulent mixed layer appears to be reached at 11 ms.

In order to make apples to apples comparison between the 3D LES simulations and the experiment, the central plane of the 3D LES domain was used to collect all data for comparison with experiment. For example, the 2D slices of heavy gas volume fraction, $Y_H$, taken from a 3D data set exemplified by Figure \ref{fig:les_3d}, were used to compute the integral mixed-width ($W$) for experimental comparison. 

An example of the temporal evolution of 2D slices are shown in comparison to experimental Mie scattering images at identical development times in Figure \ref{fig:les_2d}. The IC immediately before blast impact is shown as the far left image for the LES and experiment. One immediately notices the more diffuse nature of the simulation compared to the experiment. This may be primarily due to the numerical diffusion acting at scales below the grid resolution. In the 3D domain the limiting resolution is in the horizontal and depth directions. Because the data was taken on 2D slices of the 3D domain, the depth resolution is essentially irrelevant. In the horizontal direction the 27 LES runs were performed with 512 grid points across, giving a grid spacing of about 1 mm. In contrast to the experimental resolution ($\approx$ 0.1 mm between image pixels) this is quite coarse, so it should be kept in mind when making comparisons.

During the first several ms of the developments shown in Figure \ref{fig:les_2d} the perturbations grow similarly. A range of perturbation sizes can be seen in both LES and experiment, although more small perturbation features can be seen in the LES. The beginnings of spike tip roll-up, or KHI development, can be seen in each row as early as 2 ms. Another feature of note is a vortex pinch-off event observed in both. In the LES: the top of the large spike seen at the bottom of the LES image is starting to separate from the main spike body at 4 ms and becomes detached at 6 ms. A similar event is observed in the experiments if one tracks the second large spike down from the top of the experimental images. The tip of the vortex begins to separate at 3 ms and is seen to be clearly separated by 6 ms. The fate of this feature in the experiment is different from that of the LES in that it breaks up into much finer scales than can be seen in the LES. This is generally true for all features in the LES versus experiment, as the finer structure in experiment indicate the higher Reynolds number of the flow and the limited resolution of the LES on those scales. 

\begin{figure*}[ht!]
\centering
\includegraphics[scale=0.51]{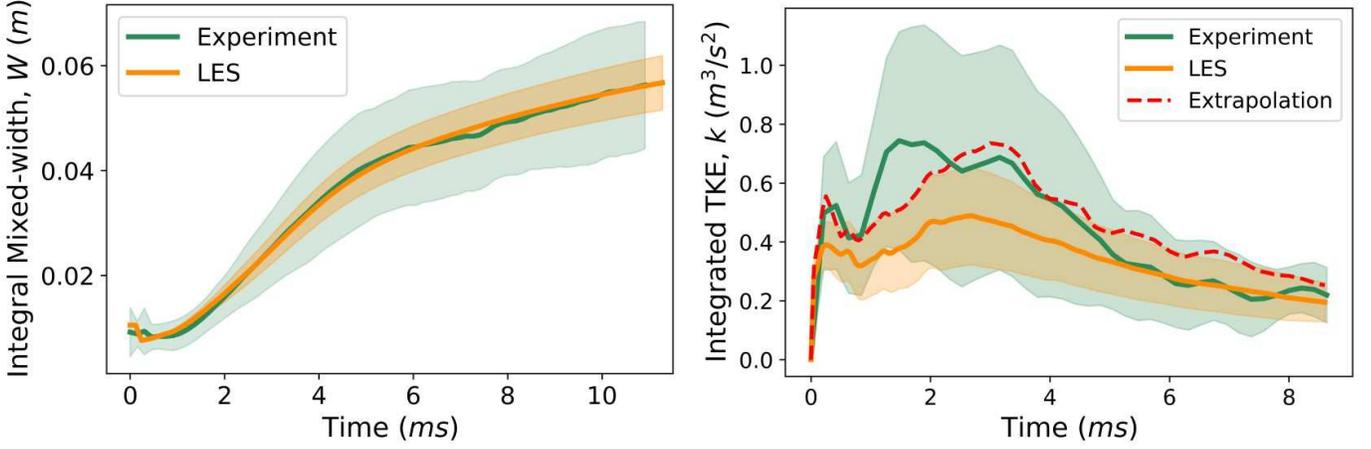}
\caption{\label{fig:les_comp} 3D LES Model: The ensemble comparison of a) integral mixed-width ($W$) and b) integrated TKE ($k$). The LES ensemble data was found using the first IC characterization method. The mean of each ensemble is shown as the bold line, while the spread of results in each ensemble is represented by the shaded region which indicates $\pm$ one standard deviation from the mean. The red dashed line in the TKE plot the Richardson extrapolated value. This gives the estimated magnitude of the TKE had the LES resolution matched that of experiment.}
\end{figure*}

One can also note the more frequent occurrence of secondary instability in the experiment as compared to the LES images. This could be due to the fact that the LES is not resolving small scale perturbations that may be seeding these secondary instabilities. A very similar phenomenon was observed in \citet{Morgan2018c} when comparing Mie Scattering experiments with Miranda LES results. The minimum interface thickness (essentially the grid spacing) necessitated by the simulations may also inhibit the secondary instabilities by effectively smearing out the gradients needed for them to develop. The disparity between the experimental scales of motion and those of the simulation is simple (at least notionally) to reduce and one need only to increase the spatial resolution of the computational mesh.  This can become cost prohibitive rather quickly and as we will show later, was not needed to capture the desired QOIs.

\begin{figure*}[t!]
\centering
\includegraphics[scale=0.51]{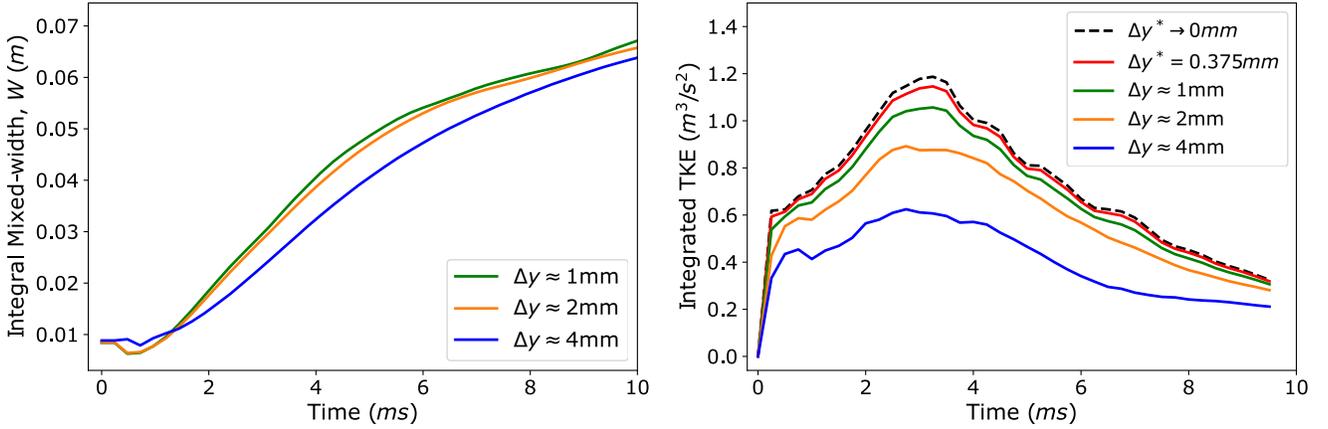}
\caption{\label{fig:les_convg} 3D LES model: The comparison of a) $W$ and b) integrated TKE for LES runs with the exact same IC, but different grid resolution. Because the variables are spanwise averaged, the resolution in the vertical direction is shown. The right panel (b) also shows the results of a Richardson extrapolation used to estimate the increase in $k$ at a resolution equal to that of the experimental PIV (red line). An extrapolation to infinite resolution is also shown as the dashed black line. All extrapolated profiles are indicated with $\Delta y^*$. The resolution of the LES ensembles is equivalent to the  $\Delta y$=2mm case shown here.}
\end{figure*}

The ensemble data comparison between the LES (using the first IC characterization method) and experiment for $W$ is shown in Figure \ref{fig:les_comp}a. $W$ is found using Equation \ref{eq:w} in an identical manner as the experiment. To make a direct comparison between the experimental 2D $W(t)$ and 3D LES results, the central plane of the 3D domain was sampled (as if it were the laser sheet in the experimental set-up), and the procedure matching that of the experiment was followed, which was outlined previously in Section \ref{sec:rans_val}. In this way the LES results can be thought of as 2D slices of a 3D data set, in the same way that the experimental data is taken from a 2D laser sheet that is sampling what is at best a quasi-2D flow field. Figure \ref{fig:les_comp}a shows that the $W(t)$ ensemble means of LES and the experiment show excellent agreement, with only a slight deviation at late times. There is also evidence of small disagreement immediately after passage of the BW, as the experiment shows a slightly more prolonged  exponential (linear) growth phase than that of the LES.

\begin{figure*}
\includegraphics[scale=0.59]{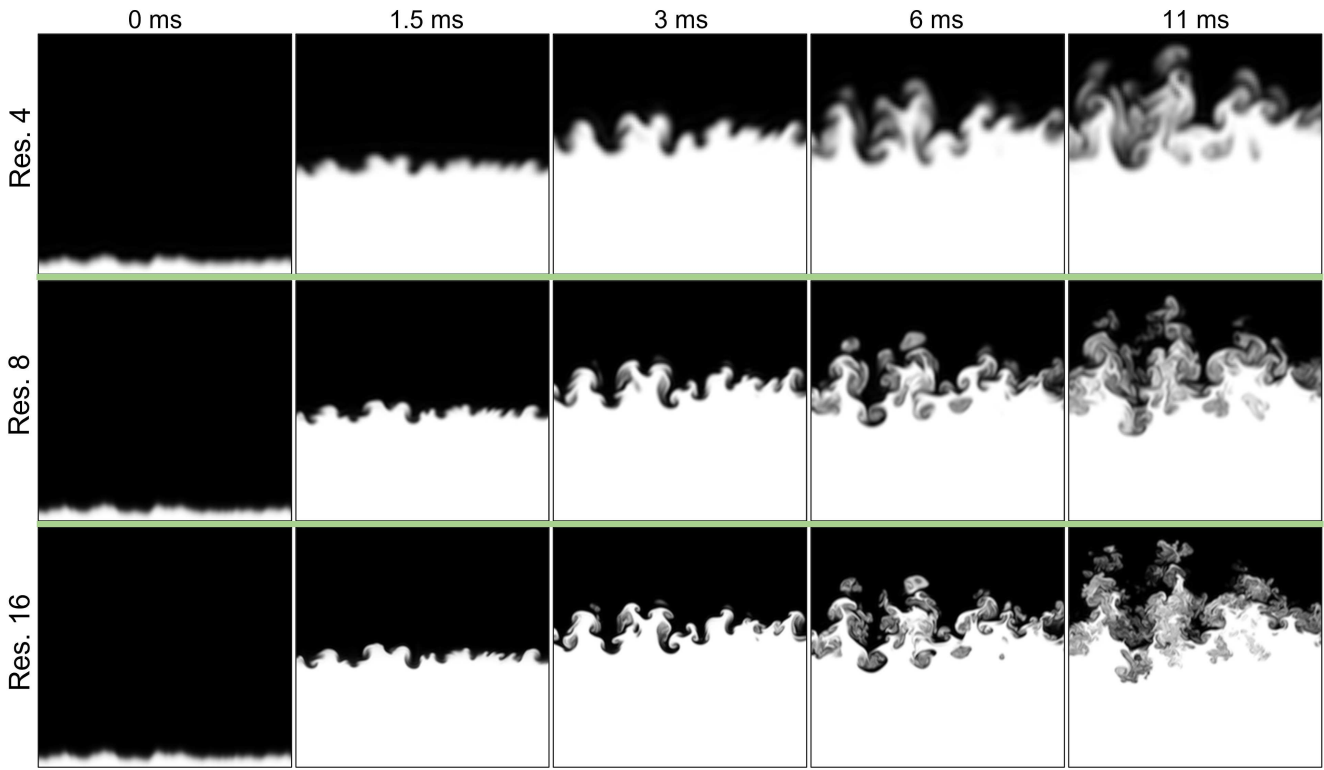}
\caption{\label{fig:convg_comp} 3D LES model: A comparison of the $Y_H$ fields for the exact same IC, but at different resolutions. The Res8 case corresponds to the resolution at which the LES ensembles were produced. Res4, Res8, and Res16 correspond to vertical resolutions ($\Delta y$) of 4mm, 2mm, and 1mm, respectively. Time proceeds from left to right. }
\end{figure*}

However, the LES does quite poorly in capturing the spread of the experimental $W(t)$ ensemble, as is shown by the shaded regions which show $\pm$ one ensemble standard deviation from the mean. There could be multiple reasons for this behavior, one being that this first IC characterization method does not properly capture the amount of variance present in the experimental ICs. This would lead to LES solutions more tightly packed about the mean. Another reason could be due to the inherent non-linearity in the system, which causes heightened sensitivity to initial conditions. Small differences in the synthetic versus experimental ICs could lead to large changes in the ensemble spread. Or, if the ICs are indeed adequately characterized by the synthetic ICs, the LES may not be replicating the high degree of non-linearity in the experimental system. This can be observed to some degree in the IC images of Figure \ref{fig:les_2d}, where the experimental image shows a much higher-degree of non-linearity in the wisps, sheets, and small vortices that are not adequately captured by the LES image. 

The LES results using the first IC characterization method are also compared to the experimental velocity (PIV) data. To make the most direct comparison to experiment, the central slice of the the 3D LES data was again taken to acquire a 2D LES velocity field in the middle of the domain. The velocity field was then mapped to a grid with a vector resolution matching that of the experiment (0.375mm/vector). The ``window" on this interpolated grid was also matched such that the LES and experiment would have the same field of view. The resulting velocity fields can then be directly compared and processed. Thus, as opposed to the RANS validation, the LES validation can directly compare the $k$ produced by experiments to that of LES.

\begin{figure*}[ht!]
\centering
\includegraphics[scale=0.52]{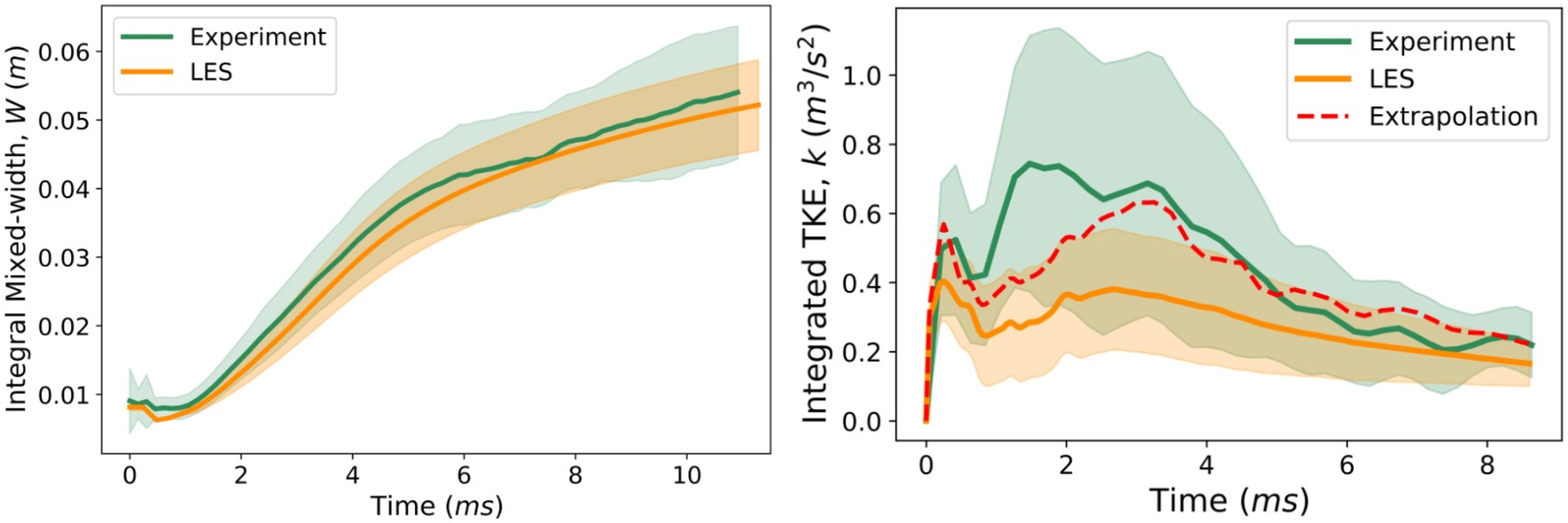}
\caption{\label{fig:les_comp2} 3D LES Model: The ensemble comparison of a) integral mixed-width ($W$) and b) integrated TKE ($k$). The LES ensemble data was found using the second IC characterization method. The mean of each ensemble is shown as the bold line, while the spread of results in each ensemble is represented by the shaded region which indicates $\pm$ one standard deviation from the mean. Te red dashed line in the TKE plot the Richardson extrapolated value. This gives the estimated magnitude of the TKE had the LES resolution matched that of experiment.}
\end{figure*}

LES velocity fluctuations were used to obtain a 2D field of $k$, which was then spanwise averaged to obtain a 1D $k$ profile for each time step in the evolution. The spanwise $k$ was then integrated in the stream-wise (vertical) direction to obtain integrated TKE (units $m^3/s^2$) for each time. This was repeated for each of the runs in the ensemble to get the mean integrated $k$ development in time. The comparison of this quantity for LES and experiment is shown in Figure  \ref{fig:les_comp}b.  Note that in this approach the $w'$ component of velocity for the LES was approximated to be $u'$, just as is done for the experiment, despite the fact that the 3D LES data produces actual values of $w'$. Thus the $k$ variable produced by experiment is reproduced exactly the same in the LES data. Note that this $k$ variable is not  technically considered turbulent kinetic energy, but is just an approximation of that variable. 

The comparison of $k$ in Figure \ref{fig:les_comp}b shows that the LES ensemble mean does not capture that of the experiment as well as it did for $W$. This is likely partially due to grid dependent effects. Because TKE is a second-order statistic, it is more sensitive to grid resolution than the 0th order $W$. Thus, it requires higher resolution to show grid convergence, whereas the LES results for $W$ may have already converged. One would expect that running the LES ensemble at a higher resolution would bump the integrated TKE values up noticeably.  Unfortunately, limited computational resources prevented the authors from performing another 27 case ensemble of 3D LES simulations at double the resolution, which equates to 16$\times$ the computational expense.

Nevertheless, one can explore the effect of grid resolution on a specific LES run. This is show in Figure \ref{fig:les_convg} where the affect of increasing grid resolution can be seen for $W$ and $k$. As can be seen in Figure \ref{fig:les_convg}, increasing the resolution has a much smaller effect on $W$ (a-left panel) than it does for $k$ (b-right panel).  For $W$, the results show that the resolution used for the LES ensembles ($\Delta y$ = 2mm) produces results essentially the same as those which are twice as resolved. This indicates that for $W$, the ensemble results are nearly grid independent and should not change for a finer resolution. 

However, for $k$ a doubling of resolution from that of the ensemble results in a notable increase in integrated TKE, as seen in Figure \ref{fig:les_convg}b.  This means that the TKE data produced in the LES ensemble has not converged. Therefore, increases in resolution should indeed push up the magnitude of $k$ seen in Figure \ref{fig:les_comp}b. In fact, even a doubling of the resolution would still not achieve the vector resolution of the PIV experiments ($\Delta y$ = 0.375 mm), or grid independence. However, using Richardson extrapolation one can estimate the magnitude increase achieved if the experimental PIV resolution were to be reached in the simulation. The results of this extrapolation (indicated as $\Delta y^*$) are shown as the red line in Figure \ref{fig:les_convg}b. The amount of increase expected to occur when the $k$ data is completely converged is shown as the black-dashed line in Figure \ref{fig:les_convg}, as this is the limit of infinite resolution. The proximity of the extrapolated red and dashed-black lines indicate that the experimental PIV is very near to capturing all of the relevant turbulent length scales. Additionally, an estimate of the LES ensemble mean, had it been performed with a resolution matching the experimental PIV data, is also shown in Figure \ref{fig:les_comp}b and Figure \ref{fig:les_comp2}b. One can see that this extrapolation improves the magnitude agreement at early times in Figure \ref{fig:les_comp}b (1st IC method), while generally improving agreement at all times in Figure \ref{fig:les_comp2}b (2nd IC method).

The effect of increased resolution can also be qualitatively examined by investigating the changes to the $Y_H$ field in the central plane of the 3D LES domain. This is shown in Figure \ref{fig:convg_comp}. As can be seen, there is a small but distinguishable difference between the three cases at all times. While the difference is more muted at the IC and immediately after BW interaction (1.5ms), one can see that the interface increasingly appears more ``crisp" with higher resolution. This is because the effective density gradient across the interface is becoming steeper with increase in resolution. In turn, the finer density gradients result in more vorticity deposition. This vorticity is deposited at the scale at which the pressure/density gradients exist. As a result, the finer density scales present at higher resolution create finer scales of baroclinic torque,  but not large scales of motion that would alter the mean flow. This is why, on the largest scales, the runs look so qualitatively similar despite the increase in resolution. However, this increase of resolution-dependent vorticity deposition at the small scales is likely a primary contributor to why the relatively under-resolved LES ensembles result in lower TKE magnitudes when compared to experiments (Figures \ref{fig:les_comp}b and \ref{fig:les_comp2}b). 

During later times, the difference in resolution manifests itself as the appearance and generation of these smaller scales. For instance, at 11 ms the Res16 case shows a range of scales that look much more characteristic of a turbulent state than the other two resolutions. These fine scale structures also serve to slightly change the shape of larger structures between the three cases. In general, an increase in resolution seems to allow  for more ``jagged" coherent structures, while the lower resolution effectively smooths out features into rounder shapes.

Finally, in order to test the sensitivity of the LES results to the characterization of ICs, a second IC approach was used. This was done to determine whether the first IC characterization method was leading to the smaller LES ensemble variance observed in Figure \ref{fig:les_comp}a. In the second IC characterization method, the IC images from the 27 experimental runs were used instead of the 300 image ensemble of ``typical ICs." The IC shape for each of the 27 runs was then essentially fed into the simulation so that each LES run had a counterpart experimental run with the same IC shape. For instance, the experimental IC shape (examples of which are in the bottom left of Figure \ref{fig:les_ic}) for each of the runs was directly input into the LES simulation. The distributions obtained in the first IC characterization scheme were used to extrapolate the IC shape to the grid boundaries in the horizontal (outside the viewing window), and into the plane. This created a 3D perturbation for which the central slice exactly matched an experimental perturbation.

The integral mixed-width results of the LES simulations using this second IC scheme are shown in Figure \ref{fig:les_comp2}a. As can be seen, feeding the IC from an experimental run directly into the LES largely increases the spread of possible results. While the mean of the $W(t)$ ensemble does not match as well as the previous approach, it is still in very good agreement. More impressively, the variance of the ensemble matches much closer than the previous results. Concerning the comparison of ensemble $k$ data shown in Figure \ref{fig:les_comp2}b, the 2nd IC characterization does little to improve the disagreement between experiments and simulation. While the Richardson extrapolated estimate arguably matches the experimental mean better than in Figure \ref{fig:les_comp}b, the general behavior is the same for both IC characterizations. This may indicate that even for the 2nd IC characterization, there is still information missing which may influence TKE development in the experiment. 

These results reinforce previous findings that show how crucial IC characterization is for LES. Despite having described the experimental ICs in a way that fully described them statistically, and where the ensemble mean was in excellent agreement, the ensemble variance was largely under predicted. In the second IC characterization method, using the exact ICs from each of the 27 experimental realizations almost doubled the amount of variance present in the LES solutions, while keeping the ensemble mean nearly the same. This is despite the fact that these same 27 ICs can be described statistically by the first IC method. Indeed, due to the non-linearity of the BDI, experiments and simulations of it are highly sensitive to ICs. Therefore for best results the LES ICs should be matched \emph{exactly} as possible in order to replicate experimental behavior.

\section{Discussion} \label{sec:discuss}

\begin{figure*}[t!]
\centering
\includegraphics[scale=0.53]{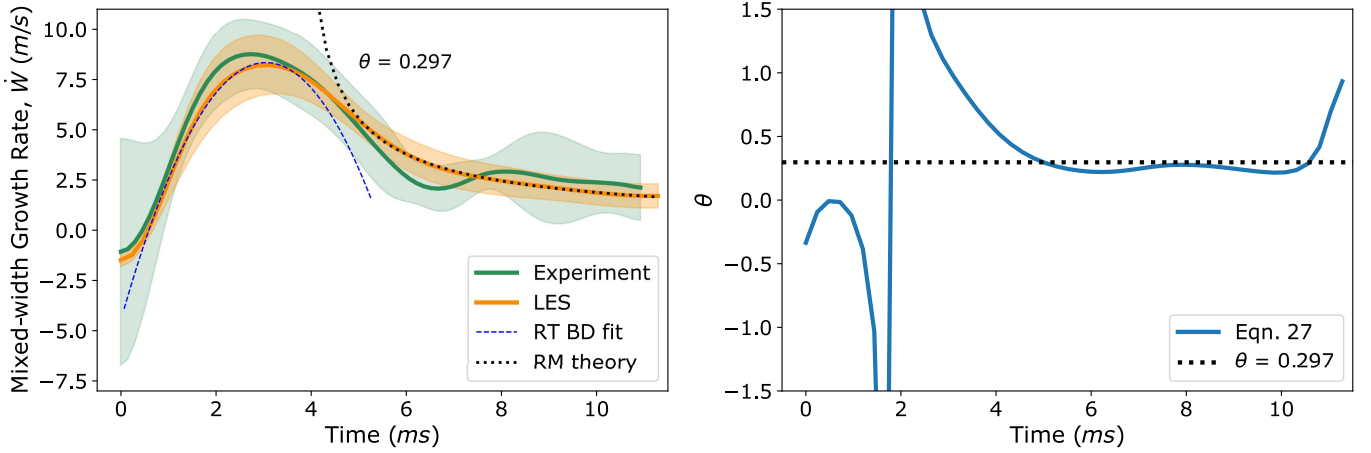}
\caption{\label{fig:les_mix_dot} 3D LES model: a) The ensemble comparison of the growth rate of $W$ for the second IC characterization method. The mean of the ensemble is shown as the bold line, while the spread of results in each ensemble is represented by the shaded region which indicates $\pm$ one standard deviation from the mean. Both the experimental and LES data are identically low-pass filtered prior to numerical differentiation. The linear+quadratic and power-law fits shown as dashed/dotted lines are applied to the LES data. b) The temporal development of $\theta$ (using LES data) according to Equation \ref{eq:theta_t} compared to the value found from the power-law fit. }
\end{figure*}

This work has shown that, generally, the LES and RANS models used in this study can reproduce experimental BDI data. Notably, the validation hierarchy is novel in that it directly connects three experimental data-sets with simulation in a staged approach.  A fully coupled validation where all uncertainties are considered simultaneously may be more complete in accounting for complex correlations, but becomes intractable when considering the cost of full 3D LES. In addition, though we found the expected mean of certain model parameters via GP regression, the distribution of these uncertain parameters was not explored but could be in future work via Markov-Chain Monte-Carlo using the GP as a surrogate.  This would allow model and experimental uncertainties to be fully fed forward into the QOI for turbulent mixing.

While different QOIs have displayed varying levels of success for each model, both RANS and LES seem to have captured at least 0th order BDI behavior. However, as expected, RANS cannot capture the turbulent transition process and therefore flows that are strongly dependent on transition will remain difficult to capture.  Indeed, RANS solutions can be calibrated to match data at certain points in time or certain regions of the flow where the approximation of fully developed flow is appropriate.  In the present study of BDI, a notable increase outside the normal range of the model coefficient ($\theta$) was required to approach good experimental agreement.  The reported $\theta$ value is in the upper range of reported  values in the RMI literature. However, as the experiments provided significance departures in geometry and instability dynamics as compared to typical RMI flows, this is unsurprising. Indeed, departure from canonical flows and geometries will likely require further investigation of RANS model coefficients in order to achieve favorable results.  

Comparably, while the 3D LES model was able to capture the 0th order mixing behavior of the BDI quite well, it was too under-resolved to replicate high-order velocity statistics. Further, even on the 0th order, it was shown to be sensitive to the method by which experimental ICs were reproduced. This observation underscores both the high non-linearity of instabilities such as BDI, but also the importance of IC characterization in LES studies. 

The run-to-run variation of the BDI experimental mixing data was large and therefore necessitated a large ensemble to reasonably bound the mean behavior. This variability is expected given the strong dependence and persistence of the ICs on the flow which is in large part due to the dominant two-dimensional flow of the facility. That is, only the finer scales become turbulent at this Atwood number. Likewise, it's reasonable to expect that simulations capturing the same narrow 3D extent would exhibit the same strong variability to ICs.  This was directly observed in the present study and shows a strong need (if not requirement) for instability driven mixing which depends largely on their initial conditions to be conducted as ensemble studies.  Otherwise, the utility of individual results will be limited to the narrow scope of those particular ICs \cite{Thornber2017}.  To the authors knowledge, the level of comparative rigor exercised in this study for shock driven mixing (where ICs were point-wise matched, post-processing was identical, ensembles were used, etc.) is unprecedented. 

The LES and experimental data can be used further to investigate the unique mixing evolution of the BDI. Three distinct mixing regimes can be identified by looking at the integral mixed-width growth-rate, $\dot{W}$, for both LES and experiment seen in Figure \ref{fig:les_mix_dot}a. The growth rate matches in both magnitude and general shape, where large changes in growth rate (regimes) are seen to occur at the same time in LES and experiment. For instance, the initial negative growth-rate at the earliest of times displays the ability of the LES to capture the phase-inversion which the BDI undergoes (due to the BW passing from higher to lower density fluid) \cite{Musci2020}. This inversion is challenging for analytic and theoretical models to capture correctly, so it is encouraging to see it replicated so well by the LES. The inflection of the growth-rate curve is also captured well by the LES as the rarefaction portion of the BW has been observed to stop interacting with the interface at approximately 3 ms. This is evident in Figure \ref{fig:les_mix_dot}a, as the 3ms point marks the inflection point of the mixed width growth-rate. 

Due to the combined nature of the instability (RMI followed by RTI), the growth-rate illustrates elements of both instabilities' traditional behavior. While similar to RMI in that a finite amount of vorticity is deposited during BW interaction, the addition of RTI prolongs that deposition duration. The signature of the RTI is seen in the growth rate of the mixing layer, where a prolonged, nearly linear, increase in mixing rate is observed during early times. This is expected as the instantaneous RMI serves to essentially pass the RTI an interface that is already growing into the non-linear regime. This should lead to enhanced mixing as compared to traditional RMI flows, which is certainly observed here. However, the divergent nature of the geometry should have an opposite effect. That is, the movement of the interface up into regions of the chamber with more space should serve to decrease the mixed width growth as the interface is forced to spread laterally. Finally, there are contributions to the mixing growth due to the prolonged compressibility of the flow field, caused by the sustained interaction of the rarefaction portion of the BW and the interface. Considering all these complications present in the BDI, it is impressive that the LES is able to so closely track its mixing growth. The LES model can be said to capture the large-scale/fundamental physics of the problem and can be more confidently used in future problems like this. 

Interestingly, Figure \ref{fig:les_mix_dot}a clearly shows three regimes of different growth-rate behavior: 1) RTI fueled growth rate increase, 2) growth-rate inflection from increasing to decreasing, and 3) RMI like growth-rate decay. In the first regime at early times, the growth-rate increases linearly.  Linear growth-rate increases are predicted by self similar RTI theory (derivative of Equation \ref{eq:r_rt}), however this is likely not applicable in the current scenario. First, it is certainly not in the self-similar growth regime, as the early images of the BDI show (Figures \ref{fig:les_2d} and \ref{fig:convg_comp}). Further, this regime of the BDI is exceptionally dynamic due to the transition from a linear to non-linear instability, large divergence effects as the interface moves up the chamber, rapid order-of-magnitude decreases in the deceleration magnitude caused by the BW, and peak compressibility effects due to the passing rarefaction. None of these complications were taken into account in the derivation of Equation \ref{eq:r_rt}. Indeed, for all the complications mentioned (divergence, compressibility, transitioning, large deceleration) it is unreasonable to expect that any theoretical estimate of RTI growth (linear or non-linear) will be able correctly predict BDI behavior in this regime. This highlights the necessity of using validated simulations to gain a better understanding in this regime.  

However, an attempt to create an ad-hoc fit originating from first principles may still be informative. If one instead assumes the mixed-width is governed by a balance of buoyancy and drag forces as has been done for many previous RTI flows \cite{Oron2001,Layzer1955, Grea2013} a more applicable form of an equation fit may be obtained. By starting with these classic buoyancy drag equations, neglecting the drag term (due to large deceleration in the regime of interest), linearizing the deceleration $g(t)$ such that 

\begin{equation}\label{eq:g_o}
g(t) \approx g_o + \frac{dg_o}{dt}t \;,
\end{equation}

\noindent and then integrating leads to

\begin{equation}\label{eq:mod_bd}
\dot{W} =  C_1 A g_o t + \frac{1}{2} A \frac{dg_o}{dt}t^2 \, C_2 + \dot{W}_o \; .
\end{equation}

\noindent Here $g_o$ is the mean deceleration over the first millisecond, and $\frac{dg_o}{dt}$ is the mean deceleration rate over that same time. The constants $C_1$ and $C_2$ are fitting parameters and $\dot{W}_o$ can be estimated using linear RMI theory:  $\dot{W}_o = W_o A \, k \, \Delta V$ \cite{Richtmyer1960}. This fit ('RT BD fit') is applied to the LES data in the first and second regime, and one can see it captures the behavior well in Figure \ref{fig:les_mix_dot}a. The linearization of the driving deceleration allows for the combination of a linear and quadratic fit, which appears to match the flow behavior. While by no means a predictive model, it could perhaps serve as a step in that direction.


In the second regime, once the BW has stopped interacting with the interface, vorticity deposition should cease and the growth rate should transition from increasing to decreasing. This regime is governed by neither traditional RMI or RTI and is likely distinct to the BDI. 

Once the growth-rate peaks and begins to decrease, the BDI enters the third regime. In this regime, one expects the growth-rate to decay according to traditional RMI theory (due to the \emph{finite} deposition of vorticity in both RMI and BDI). This decay can be expressed as $W = B \, \theta (t -t_o)^{\theta -1}$, where $B$ is a dimensional fitting constant and $t_o$ is the time at which the BW stops interacting with the interface \cite{Thornber2017}. A power-law according to this equation is fit to the LES data in Figure \ref{fig:les_mix_dot}a, and the best-fit value for $\theta$ is shown on the plot. Observing this relationship,one can see that RMI theory seeks the behavior of decaying, anisotropic, and homogeneous turbulence. This is certainly applicable for the BDI mixing layer during mid-late times (Figures \ref{fig:les_2d} and \ref{fig:convg_comp}) where the RMI fit is applied. One can also get an idea of how the RMI decay exponent, $\theta$, changes in time by following methods used in \citet{Thornber2017}. Namely, a time varying $\theta$ value is estimated using:

\begin{equation}\label{eq:theta_t}
\theta^{-1} = \left( 1- \ddot{W} \, W \, / \: \dot{W}^2 \right) \;.
\end{equation}

\noindent Applying this relationship to the LES data results in the temporal $\theta$ behavior shown in Figure \ref{fig:les_mix_dot}b. The horizontal dashed line indicates the $\theta$ value found using the power-law fit, which is seen to agree quite well with the asymptotic value predicted by Equation \ref{eq:theta_t}. The spikes seen at early-mid times are due to the BDI behavior in the first and second regimes, where Equation \ref{eq:theta_t} obviously does not apply. The mid-late time behavior displayed in Figure \ref{fig:les_mix_dot}b further indicates that the BDI decays very similarly to the RMI. Indeed the $\theta$ values found using the power-law and Equation \ref{eq:theta_t} fall within the range reported in literature \cite{Zhou2017b, Thornber2017}.

\section{Conclusion} \label{sec:conclusion}

This article presented a hierarchical validation scheme that used ensembles of experimental data to compare QOIs with three different models. First, a 2D Euler model was used to recreate the experimental facility and its driving physics. Boundary and initial conditions were successfully modeled and a drag-based loss model was added to account for unmeasured losses in the experimental facility. GP optimization showed the tuned Euler model was relatively insensitive to changes in model parameters. 

The validated Euler model was then used as a basis for the implementation of a 2D RANS and 3D LES model, which were validated against ensembles of mixing data from separate experimental campaigns. It was necessary to use three model parameters for optimizing the RANS model, which showed good agreement in mixing statistics once the flow approached turbulence. The optimized model also showed that the rate of energy dissipation in the RANS model matched that of the experiment, despite large differences in TKE magnitude. To validate the LES model, we generated a simulation ensemble of equal size to the experimental data ensemble. This allowed for the comparison of mean and variance trends for the mixing QOI. While, the validated LES model had no model parameters to optimize it was shown to be sensitive to the method used for the characterization of ICs. Directly inputting experimental ICs into the LES model led to much better overall agreement between the data ensembles for both mixing width and its growth rate. The LES model was shown to capture all of the relevant physics present in the BDI experiments. The LES also helped to elucidate information about the three mixing regimes observed in the BDI. Both similarities and differences with traditional RTI and RMI were observed. Primarily, the BDI was shown to decay according to RMI theory at late times. 

Further work with these data sets can be pursued by investigating further comparison between LES and RANS simulations for QOIs that are not available experimentally. This could also help inform more general comparison between the RANS and LES models themselves, beyond just the QOI outputs. Additionally, a more formal uncertainty quantification of model parameters could be completed. For example, acquiring distributions of parameter uncertainty as opposed to just means. Further, much more work can be done to elucidate the complex behavior of the first mixing regime. Validated simulations can be compared to a range of linear and non-linear theories, or analytical models, that have been developed to address different complications in the traditional RTI.   

The resulting validated digital twin can be used in numerous direction for future work. It can be used to inform design modifications to the experimental facility, or for experimental campaign planning when expanding the tested parameter space. The parameter space of BW strength and Atwood number can be further, and more easily, investigated using the validated RANS and LES models to investigate different BDI regimes. Further the results of these investigations could allow for the creation of GPs that replicate the RANS and LES models themselves. This would allow for a much more computationally efficient investigation of very large areas in the parameter space of the BDI. 

\section{Acknowledgment} \label{sec:acknow}
The authors wish to thank Dr. Brandon Morgan for his consultation on using the RANSbox and for his invaluable insight concerning the interpretation of RANS results. Thanks are also in order for Erin Wilson regarding her immense contributions to the aesthetic of plots and figures in this article. The authors also wish to thank Ben's van, which facilitated much of this work.

The first author is also incredibly thankful for the funding and opportunities provided by the DOE NNSA Stewardship Science Graduate Fellowship (SSGF) and the LLNL Defense Science and Technology Fellowship (DSTI). In addition, we acknowledge support from the DOE Early Career Award and Office of Fusion Energy Science in the form of grant DE‐SC0016181. Finally, this work was performed under the auspices of the U.S. Department of Energy by Lawrence Livermore National Laboratory under Contract DE-AC52-07NA27344.


\section{Nomenclature} \label{sec:nom}

BC -- Boundary condition

BDI -- Blast-Driven Instability 

BW -- Blast Wave

DCT -- Discrete Cosine Transform

DPT -- Dynamic pressure transducer

GP -- Gaussian Process

IC -- Initial Condition

ICF -- Inertial Confinement Fusion

KHI/KH - Kelvin-Helmholtz Instability

LES -- Large Eddy Simulation

LLNL - Lawrence Livermore National Lab

PIV -- Particle Image Velocimetry

QOI -- Quantities of interest

RANS -- Reynolds Averaged Navier Stokes

RMI/RM -- Richtmyer-Meshkov Instability

RP80 -- Detonator with 0.2g high explosive

RTI/RT -- Rayleigh-Taylor Instability

SN -- Supernova

TKE -- Turbulent Kinetic Energy

$A$ -- Atwood number - $\frac{\rho_{h} - \rho_{l}}{\rho_{h} + \rho_{l}}$

$a$ -- 1/2 Peak-to-Valley amplitude or Mixed-width

$C_o$ -- Euler model drag coefficient

$E_o$ -- Euler model initial pill energy

$\dot{a}$ -- Mixed-width growth rate

$g(t)$ -- Interface acceleration

$\Gamma$ -- Model dependent source/sink terms

$K$ -- RANS model turbulent kinetic energy

$k$ -- Experimental estimate of turbulent kinetic energy

$\kappa$ -- Wave-number

$L$ -- RANS model turbulent length scale

$\lambda$ -- Wavelength

$M$ -- Mach number

$Re$ -- Reynolds number

$\theta$ -- RM self-similar growth exponent

$W$ -- Integral mixed-width

$Y_H$ -- Heavy gas mass fraction



%

\end{document}